\documentclass[a4paper,fleqn,usenatbib,useAMS]{mnras}

\usepackage{graphicx}	% Including Figure files
\usepackage{amsmath}	% Advanced maths commands
\usepackage{amssymb}	% Extra maths symbols
\usepackage{multicol}        % Multi-column entries in Tables
\usepackage{bm}	% Bold maths symbols, including upright Greek
\usepackage{pdflscape}	% Landscape pages
\usepackage{subfigure}
\usepackage{longtable}
\usepackage{array}
\usepackage{booktabs}

\usepackage[T1]{fontenc}
\usepackage{ae,aecompl}

\title[Cocoon breakout]{Cocoon breakout and escape from the ejecta of neutron star mergers}

\author[Hamidani \& Ioka]{Hamid Hamidani$^{1,2}$\thanks{E-mail: hamidani.hamid@yukawa.kyoto-u.ac.jp} and Kunihito Ioka$^1$
\\
$^{1}$ Yukawa Institute for Theoretical Physics, Kyoto University, Kyoto 606-8502, Japan\\
$^{2}$ Astronomical Institute, Tohoku University, Aoba, Sendai 980-8578, Japan\\
}
\date{Accepted XXX. Received YYY; in original form ZZZ}

\pubyear{2022}

\begin{document}
\label{firstpage}
\pagerange{\pageref{firstpage}--\pageref{lastpage}}
\maketitle

\begin{abstract}
The cocoon is an inevitable product of a jet propagating through ambient matter, and takes a fair fraction of the jet energy. 
In short gamma-ray bursts, the ambient matter is the ejecta from the merger of neutron stars, expanding with a high velocity $\sim 0.2 c$, in contrast to the static stellar envelope in collapsars. 
Using 2D relativistic hydrodynamic simulations with the ejecta density profile as $\rho \propto r^{-2}$, we find that the expansion makes a big difference;
only 0.5--5\% of the cocoon mass escapes from (faster than) the ejecta, with an opening angle $20^{\circ}$--$30^{\circ}$, 
while it is $\sim 100\%$ and spherical in collapsars. 
We also analytically obtain the shares of mass and energies for the escaped and trapped cocoons. 
Because the mass of the escaped cocoon is small and the trapped cocoon is concealed by the ejecta and the escaped cocoon, we conclude that it is unlikely that the cocoon emission was observed as a counterpart to the gravitational wave event GW170817.
\end{abstract}

%%%%%%%%%%%%%%

\section{Introduction}
\label{sec:1}
%GW170817 discovery
In August 17th 2017, the Laser Interferometer Gravitational-Wave Observatory (LIGO) and the Virgo Consortium (LVC) detected the first gravitational wave (GW) signal from the binary neutron star (BNS) merger event, GW170817 (\citealt{2017PhRvL.119p1101A}).
About $1.7$ s after the GW signal, Fermi recorded a \textit{short} Gamma-Ray Burst (\textit{s}GRB), \textit{s}GRB 170817A  (\citealt{2017ApJ...848L..13A}). 
This ultimately confirmed the scenario of BNS mergers for \textit{s}GRBs (\citealt{1986ApJ...308L..43P}; \citealt{1986ApJ...308L..47G}; \citealt{1989Natur.340..126E}). 
It took about $10$ hours to finally localize the merger site, and an intense follow-up observation campaign across the electromagnetic (EM) spectrum followed soon after. 
Hence, a new era of multi-messenger astronomy started (\citealt{2017ApJ...848L..13A}).

% two important discoveries
In particular, this campaign enabled the discovery of the kilonova/macronova (KN hereafter), and its analysis confirmed the presence of the expanding ejecta, measured its properties (e.g., mass and velocity), and found indications for r-process nucleosynthesis (of heavy and unstable elements) within it (\citealt{Arcavi:2017vbi}; \citealt{Chornock:2017sdf}; \citealt{Coulter:2017wya}; \citealt{2017ApJ...848L..29D}; \citealt{2017Sci...358.1570D}; \citealt{Kilpatrick:2017mhz}; \citealt{2017Sci...358.1559K}; \citealt{Nicholl:2017ahq}; \citealt{Pian:2017gtc}; \citealt{Smartt:2017fuw}; \citealt{Shappee:2017zly}; \citealt{Soares-Santos:2017lru}; \citealt{2017PASJ...69..102T}; \citealt{Utsumi:2017cti}; \citealt{Valenti:2017ngx}), as previously predicted (\citealt{1998ApJ...507L..59L}; \citealt{2005astro.ph.10256K}; \citealt{2010MNRAS.406.2650M}).
This campaign was also able to find clear evidence of a relativistic jet (\citealt{2018Natur.561..355M}) viewed off-axis (also see \citealt{2018PTEP.2018d3E02I} and \citealt{2019MNRAS.487.4884I}).
These discoveries are perfectly consistent with the scenario of \textit{s}GRBs.

%the emergence of the cocoon
In this scenario, the merger of two compact objects, NS-NS (or Black Hole; BH-NS), produces a system of a compact remnant with an accretion disk.
This system is what powers the relativistic jet of \textit{s}GRBs (\citealt{1986ApJ...308L..43P}; \citealt{1986ApJ...308L..47G}; \citealt{1989Natur.340..126E});
hence, it is often refereed to as the ``central engine".
A clean relativistic jet in the line of sight of the observer is the essential ingredient to explain the prompt emission of \textit{s}GRBs.
However, after the merger, the expanding ejecta surrounds the central engine (\citealt{1999PhRvD..60j4052S}; \citealt{2000PhRvD..61f4001S}).
Therefore, in order for the \textit{s}GRBs to be observed, the jet has to make its way across the ejecta. 
While the jet is faster than the ejecta, the ejecta is much denser.
As a result of the jet-ejecta interaction, the head of the highly relativistic jet is slowed down, and a shock structure (jet head) is created (\citealt{1974MNRAS.169..395B}; \citealt{1974MNRAS.166..513S}).
During this phase, the jet outflow is continuously mixed with the ejecta, creating a hot, turbulent, and highly pressurized component in the surroundings of the jet called the ``cocoon" (\citealt{1989ApJ...345L..21B}).
Once the outer edge of ejecta is reached, both of the jet and the cocoon can escape to the outside of the ejecta, i.e., breakout.
This picture has been confirmed by numerical simulations (\citealt{2014ApJ...784L..28N}; \citealt{2014ApJ...788L...8M}).

% more to be achived >> cocoon?
Although late time observations of GW170817 have dramatically improved our theoretical understanding of \textit{s}GRBs and their environment, many questions remain unanswered, in particular related to the central engine of \textit{s}GRBs.
For instance, the nature of the remnant (BH or NS), the jet launch process, and its nature are still open questions, and would remain open without observational breakthroughs.

%cocoon interesting 
The cocoon is an interesting component in several aspects.
It is an intermediate component between the jet and the ejecta (e.g., in terms of speed and rest-mass density).
After the breakout, the cocoon immediately expands outward.
As it is expected to carry out decent energies and achieve mildly relativistic speeds, theoretically it could power a unique astrophysical transient (\citealt{2017ApJ...834...28N}; \citealt{2017ApJ...848L...6L}; \citealt{2018MNRAS.473..576G}; \citealt{2018PTEP.2018d3E02I}; etc.).
Also, monitoring the cocoon emission could theoretically help learn more about \textit{s}GRBs (e.g., the jet and the central engine) and the KN (e.g., r-process nucleosynthesis in early times).

%cocoon could hold the key
Unfortunately, it took $10$ crucial hours to localize GW170817 in the sky, and for EM observations to start;
the opportunity to detect the cocoon emission has most likely been missed (considering estimations by: \citealt{2018PTEP.2018d3E02I}; \citealt{2018MNRAS.473..576G}; \citealt{2021MNRAS.500.1772N}; \citealt{2021MNRAS.502..865K}; and others). 
However, this is expected to change in a very near future. 
With the new generation of GW detectors, multi-messenger observations of GW170817-like events are expected to be more frequent, and sky localization is expected to take much less time, presenting more opportunities for the detection of the cocoon emission, and for new breakthroughs.

%simualtions previous studies
Thanks to GW170817, interest in the subject of the sGRB jet has grown, and so did our understanding.
Most works used numerical simulations to investigate jet propagation in the expanding ejecta, the effect of the jet on the KN, the cocoon emission, etc. (\citealt{2014ApJ...784L..28N}; \citealt{2015ApJ...813...64D}; \citealt{2017ApJ...848L...6L}; \citealt{2017Sci...358.1559K}; \citealt{2018MNRAS.473..576G}; \citealt{2018ApJ...866....3D}; \citealt{2018MNRAS.479..588G}; \citealt{2018ApJ...863...58X}; \citealt{2018MNRAS.475.2971B}; \citealt{2020MNRAS.491.3192H}; \citealt{2021MNRAS.500.1772N}; \citealt{2021MNRAS.500..627H}; \citealt{2021MNRAS.500.3511G}; \citealt{2021MNRAS.502..865K}; \citealt{2021ApJ...918L...6L}; \citealt{2021MNRAS.503.4363U}; \citealt{2022MNRAS.509..903N}; and references within). 
Numerical simulations present a powerful tool to study the cocoon.
However, as most simulations ended prematurely (either due to numerical requirements or to a focus on the jet), the process of ``cocoon breakout" and the cocoon's late time evolution are yet to be studied. 

%Analyic
There has been a few attempts to analytically model the cocoon in \textit{s}GRBs, and estimate its emission.
\cite{2017ApJ...834...28N} estimated the pre-breakout cocoon in the context of collapsars (where the medium is static), and its emission.
Then, they applied their model to the context of \textit{s}GRBs giving estimates of the pre-breakout cocoon's properties, and used these properties to deduce the cocoon emission.
Using this model, \cite{2018ApJ...855..103P} claimed that GW170817 showed clear evidence for a cocoon component (assuming a cocoon mass in the order of $0.01 M_\odot$).
\cite{2018PTEP.2018d3E02I} presented another analytical estimate of the pre-cocoon mass and average velocity in \textit{s}GRB, and discussed its possible contribution to the blue KN component. 

%there has been a logical jump in the past, from pre- to emission. it should be pre- post- emission
In \textit{s}GRBs the cocoon is fundamentally different than that in collapsars (\citealt{2021MNRAS.500..627H}).
In collapsars, as the medium is static, most of the cocoon is expected to breakout; therefore, the pre-breakout cocoon is quite reliable to estimate the cocoon emission.
However, in \textit{s}GRBs, as the ejecta is initially expanding ($\sim 0.2c$; \citealt{2013PhRvD..87b4001H}; \citealt{2013ApJ...773...78B}; \citealt{2015MNRAS.448..541J}), the situation is much trickier; only a fraction of the cocoon is faster than the ejecta, and hence gets to escape from the ejecta
after the breakout (as shown later).
Almost all the cocoon is trapped and ends up being hidden by the ejecta (and the escaped cocoon), and not observable.
Hence, it is important to distinguish between the ``escaped cocoon" and the ``trapped cocoons".
Furthermore, the internal energy composition is also quite different than in collapsars (see Figure 1 in \citealt{2021MNRAS.500..627H}).
Therefore, in order to analytically estimate the cocoon emission, understanding the cocoon breakout in the expanding ejecta, and finding the late time (i.e., post-breakout) cocoon's properties is essential.
Hence, previous analytical estimates of the cocoon emission, although very reasonable in the context of collapsars (e.g., \citealt{2017ApJ...834...28N}), are not quite as reasonable in the context of \textit{s}GRBs. 

% goal and originality
Here, aiming to understand the cocoon emission in NS mergers, model it, and use it to learn more about NS mergers and their environment (e.g., central engine of \textit{s}GRBs, and r-process nucleosynthesis), we present a rigorous study of the cocoon breakout from the ejecta of NS mergers.
First, we use hydrodynamical numerical simulation, in order to understand the process of the cocoon breakout.
We follow the cocoon evolution for timescales sufficiently longer than the timescales of the jet breakout and the engine activity (for a total of $10$ s after the jet launch).
Then, we construct a fully analytic model to solve the cocoon breakout, allowing us to systematically infer the properties of the post-breakout cocoon (as a function of the parameters of the jet and the ejecta).
Finally, we evaluate our analytic model by comparing its results to numerical simulations. 

%summary
This paper is organized as follows. 
In Section \ref{sec:2}, numerical simulations of the cocoon breakout from NS merger ejecta are presented, and their results are interpreted.
In Section \ref{sec:3}, analytic modeling of the pre-breakout and post-breakout cocoon is presented.
In Section \ref{sec:4}, analytic results are compared with simulations, and discussed.
Finally, a conclusion is presented in Section \ref{sec:5}.

%%%%%%%%%
\section{Cocoon breakout in numerical simulations}
\label{sec:2}
\subsection{Setup and jet models}
\label{sec:2 setup}
We use the same numerical code as in \cite{2017MNRAS.469.2361H}, \cite{2020MNRAS.491.3192H}, and \cite{2021MNRAS.500..627H}.
The numerical procedure and the numerical setup is the same as in \cite{2021MNRAS.500..627H} [also see \citealt{2020MNRAS.491.3192H} for more technical details].
However, here we only investigate the jet propagation in \textit{s}GRB -- NS merger context (the expanding medium case, as labeled in \citealt{2021MNRAS.500..627H}).
Table \ref{tab:1} shows the representative subsample of jet models to be studied here: ``narrow", ``wide", and ``failed".

The initial density profile of the ejecta is taken as a single power-law with an index $n$.
Ideally, the density profile of the ejecta should follow a broken power-law with the inner part having an index $n\sim 2$, and $n\sim 2.5-3.5$ for the outer part (shallow in the polar region, and steep in the equatorial region; see Figure 8 in \citealt{2020MNRAS.491.3192H}).
Here, as we are mainly focused on the jet-cocoon evolution, we make two simplifications.
i) we take $n= 2$ throughout the ejecta, and ii) we take the ejecta as spherically symmetric, using the polar densities\footnote{\label{foot:1}In other words, we are subtracting the extra mass near the equatorial region, relative to the polar region. 
This would result in a reduction of the ejecta total mass by $\sim 1/5$ (for more details see footnote 8 in \citealt{2020MNRAS.491.3192H}). 
Therefore, the $0.002 M_{\odot}$ in our models is based on a fiducial total ejecta mass of $0.01 M_\odot$ (again see \citealt{2020MNRAS.491.3192H}).}.
We set the ejecta to expand homologously $r\propto v$ as suggested by numerical relativity simulations (again, see Figure 8 in \citealt{2020MNRAS.491.3192H}).
We set the maximum velocity of the ejecta as $\beta_m = \sqrt{3}/5$ (so that the average velocity of the ejecta is $\sim 0.2c$, where $c$ is the speed of light; see \citealt{2013PhRvD..87b4001H}; \citealt{2013ApJ...773...78B}; \citealt{2015MNRAS.448..541J}; or see Section 4.2 in \citealt{2020MNRAS.491.3192H}).

For simplicity, our numerical simulations do not take into account the fast tail of the ejecta (this simplification is discussed in Section \ref{sec:limits}).
On the other hand, the circumstellar medium's (CSM) density is set as $\rho_{CSM} = 10^{-10}$ g/cm$^3$.
This density is obviously much higher than what is expected for BNS merger's surrounding.
This choice has been taken to prevent numerical problems that often arise from huge density gaps.
Such high density would affect the post-breakout cocoon (and jet) at much later times (in particular in the much weaker narrow jet case).
However, since our simulations end at $t-t_0 = 10$ s, the total swept CSM mass is still negligible (i.e., relative to the mass of the escaped cocoon).

%%%%%%%%%%%%%%%%%%%%

\begin{table}
\caption{
The subsample of the simulated models and their corresponding parameters. 
From the left: 
The model name (depending on the jet type or its fate); 
the ejecta mass, assuming polar densities [these values should be multiplied by $\sim 5$ to find the total ejecta mass (to account for high densities in the equatorial region); see Section \ref{sec:2 setup} and footnote \ref{foot:1}]; 
the jet initial opening angle;  
and the engine's isotropic equivalent luminosity [$L_{iso,0} = \frac{2 L_j}{1-\cos\theta_0}\simeq \frac{4 L_j}{\theta_0^2}$] where $L_j$ is the jet true luminosity (one sided).
All the other parameters are the same for the three jet models: 
the inner radius at which the jet is injected in simulations is set as $r_0 = 1.2\times 10^{8}$ cm; 
the ejecta's initial density profile is taken as a power-law function with an index $n=2$;
the maximum velocity of the ejecta is taken as $\beta_m=\sqrt{3/25} \approx 0.345$;
the delay time between the merger time (i.e., the launch of the ejecta) and the jet launch time is taken as $t_0-t_m = 0.160$ s.}
\label{tab:1}
\begin{tabular}{l|lll}
\hline
 Jet models  & $M_{e}$ [$M_\odot$] & $\theta_0$  [deg] & $L_{iso,0}$ [erg s$^{-1}$]  \\
  \hline
  Narrow & $0.002$ & $6.8$ & $5\times10^{50}$ \\
  Wide  & $0.002$ & $18.0$ & $5\times10^{50}$ \\
  Failed & $0.010$ & $18.0$ & $1\times10^{50}$ \\
    \hline
  \hline
 \end{tabular}
\end{table}
%%%%%%%%%%%%%%%%%%%%%%%%

\subsection{The timeline}
\label{sec:timeline}
Simulations are set to start at $t=t_0$.
The jet is launched (injected) at the same time, for a duration of $t_e-t_0 = 2$ s.
The delay between the merger time and the jet launch time is set as $t_0-t_m=0.160$ s.
All simulations are run, through the jet breakout, until $t-t_0 = 10$ s (breakout times are listed in Table \ref{tab:2}).
This is considerably a much longer simulation time compared to previous studies (e.g. \citealt{2021MNRAS.500..627H}; \citealt{2018MNRAS.473..576G}) and requires a large computational domain.

The motivation behind this longer computation time is to follow the late time evolution of the cocoon, until the free expansion phase is reached and the system is fully ballistic,
i.e., interaction between the jet/cocoon/ejecta becomes negligible.
We refer to this time, the time at which the system is ballistic, as $t_1$.
This time can roughly be estimated as the time when the last jet outflow (after the engine is turned off) breaks out of the ejecta.
One can easily estimate this time as $t_1 -t_0 \sim (t_e-t_0)/(1-\beta_m) \sim 3$ s.
For the failed jet case, the cocoon breakout happens at much later times, even later than $t_e \sim 2$ s.
Although, ideally, much longer simulation time is needed, we set $t_1-t_0\sim 10$ s based on the evolution of the cocoon showing no significant changes at this time (taking into account the adiabatic expansion), indicating that late time interactions are weak. 

Hence, to summarize, assuming a merger time $t_m$, the jet is set to launch at $t_0=t_m+0.160$ s, for a duration of $2$ s, until $t=t_e$.
The breakout happens before $t_e$ for successful jets (narrow and wide here), or after $t_e$ for the failed jet case.
After the breakout the jet-cocoon continues to interact with the ejecta until $t\sim t_1$, roughly $\sim 3$ s for successful jets, and $\sim 10$ s for the failed jet case (from $t_0$).
The aim of this simulation study is to analytically model the cocoon so that simulation results at $t\sim t_1$ can be understood and systematically reproduced.

\subsection{The pre-breakout cocoon: shocked jet vs. shocked ejecta}
\label{sec:pre-breakout sim}
\begin{figure*}%[ht] 
    \vspace{4ex}
  \begin{subfigure}%[b]{0.2\linewidth}
    \centering
    \includegraphics[width=0.49\linewidth]{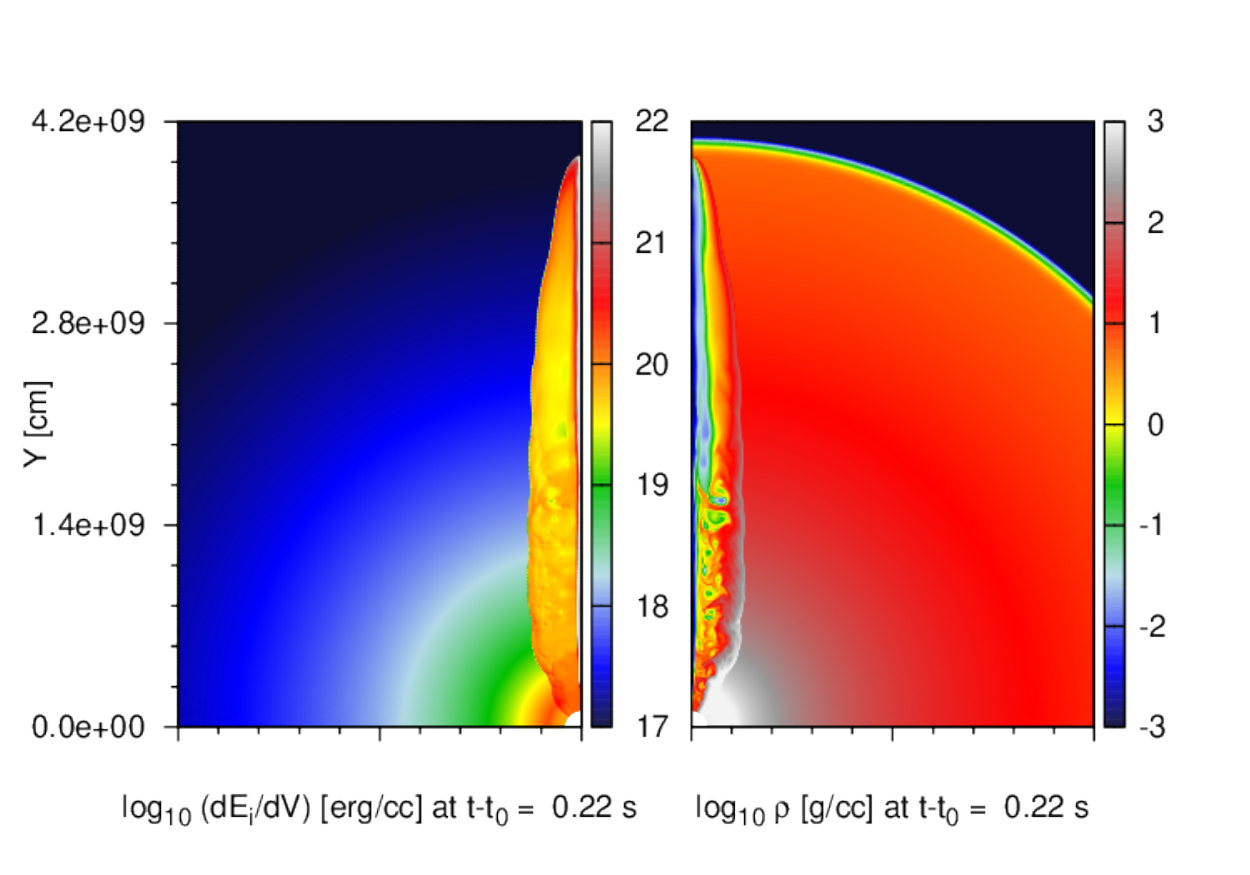}
    %\vspace{4ex}
  \end{subfigure}%% 
  \begin{subfigure}%[b]{0.2\linewidth}
    \centering
    \includegraphics[width=0.49\linewidth]{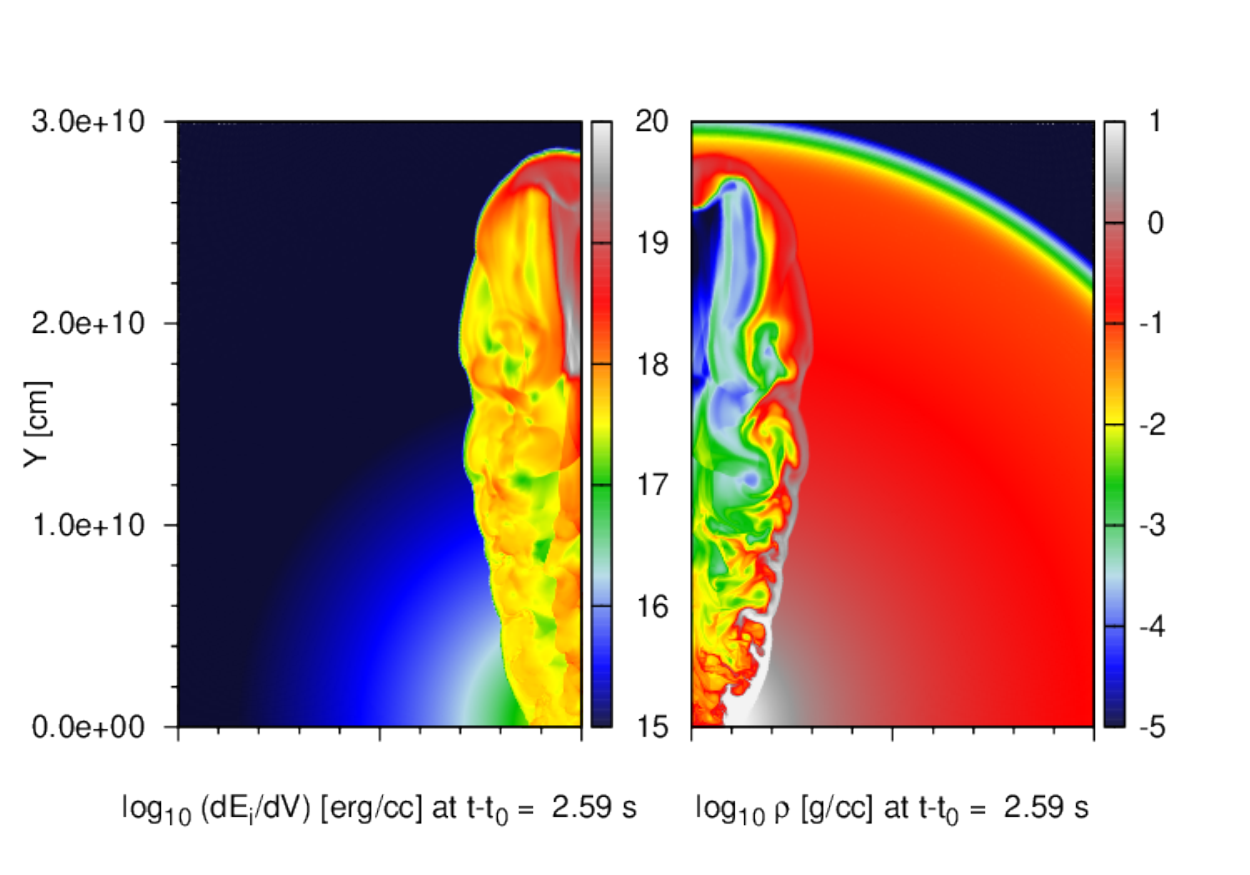} 
    %\vspace{4ex}
  \end{subfigure}
  %\vspace{4ex}
  \caption{
  Internal energy density (lab; left) and rest mass density (comoving; right) maps of the pre-breakout cocoon in numerical simulations, as it is about to break out of the ejecta. 
  The two jet models presented here are the narrow jet (left sub-figure) and failed jet (right sub-figure) [see Table \ref{tab:1}].
  }
  \label{fig:1 sim} 
\end{figure*}

Left panels in Figure \ref{fig:1 sim} show the internal energy density map at the moment of the breakout\footnote{\label{foot:labo}Unless specified, quantities are calculated in the laboratory frame.}.
Two models are shown, the narrow jet (left) and the failed jet (right), and their corresponding cocoon is clearly visible.
Previous studies have shown that for the cocoon of a collapsar jet, the internal energy density (i.e., pressure) is almost homogeneously distributed (\citealt{2011ApJ...740..100B}; \citealt{2013ApJ...777..162M}; \citealt{2018MNRAS.477.2128H}; etc.).
The same has been found for the BNS merger case, where the ejecta is expanding (see \citealt{2018MNRAS.473..576G}; \citealt{2021MNRAS.500..627H}).
Here, our simulations show that the internal energy distribution is very consistent with these previous findings. 
This feature will be essential for modeling the cocoon breakout (in Section \ref{sec:3}).

Previous work by \cite{2011ApJ...740..100B} explained in a simplified picture that, at a given time before the breakout, the cocoon can be divided into two contrasting parts: i) the inner cocoon part, consisting of shocked jet material; and ii) the outer cocoon part, consisting of shocked medium material (see Figure 1 in \citealt{2011ApJ...740..100B}).
\cite{2011ApJ...740..100B} explained that the inner cocoon part has a low mass density, and a high fraction of internal energy in its total energy;
while the outer cocoon part has a much higher mass density and a lower fraction of internal energy in its total energy.
Here, the medium (i.e., the ejecta) is expanding, compared to \cite{2011ApJ...740..100B} where the medium is static (i.e., collapsar) [more details are given in Section \ref{sec:3}].
Nevertheless, as it can be seen from  Figure \ref{fig:1 sim} (right panels; showing the rest mass density map), the same pattern for the two parts can be identified; 
despite the presence of an intermediate region where these two parts are mixed (more discussions see Section \ref{sec:breakout of the two cc parts} and \ref{sec:fmix}).
\cite{2017ApJ...834...28N} relied on these two parts for their modeling of the cocoon in collapsars.
Here too, these two parts and their properties will be decisive for modeling the cocoon breakout, and understanding the post-breakout cocoon's properties (i.e., mass, energy, and internal energy) [see Section \ref{sec:3} and Figure \ref{fig:keyA} for a general idea].

%%%%%%%%%%%
%%%%%%%%%%%%%%%%%%%%%%%%%%%%%%%%%%%%%%

\subsection{The post-breakout cocoon: the escaped vs. the trapped cocoon}
\label{sec:post-breakout sim}
We followed the cocoon for much longer times after it broke out of the ejecta, until $t-t_0=10$ s. 
In the following we will analyze its properties after the breakout.
First, let's recall that the maximum achieved Lorentz factor for a stationary fluid element can be estimated using Bernoulli equation as\footnote{Note that for the cocoon of NS mergers, the radial component of velocity is much more dominant: $\beta_r \gg \beta_\theta$.
Hence, it is safe to neglect $\beta_\theta$ giving $\beta \sim \beta_r$ (in particular at larger radii of the ejecta).}
\begin{eqnarray}
    \Gamma_{inf} \approx h \Gamma ,
    \label{eq:Bernoulli Gamma}
\end{eqnarray}
where $h$ is the enthalpy.
Using $\Gamma_{inf}$, we can estimate the maximum velocity as (see footnote \ref{foot:labo})
\begin{eqnarray}
    \beta_{inf}=\sqrt{1-1/\Gamma_{inf}^{2}} .
    \label{eq:Bernoulli Beta}
\end{eqnarray}

\begin{figure*}%[ht] 
    \vspace{4ex}
  \begin{subfigure}%[b]{0.2\linewidth}
    \centering
    \includegraphics[width=0.49\linewidth]{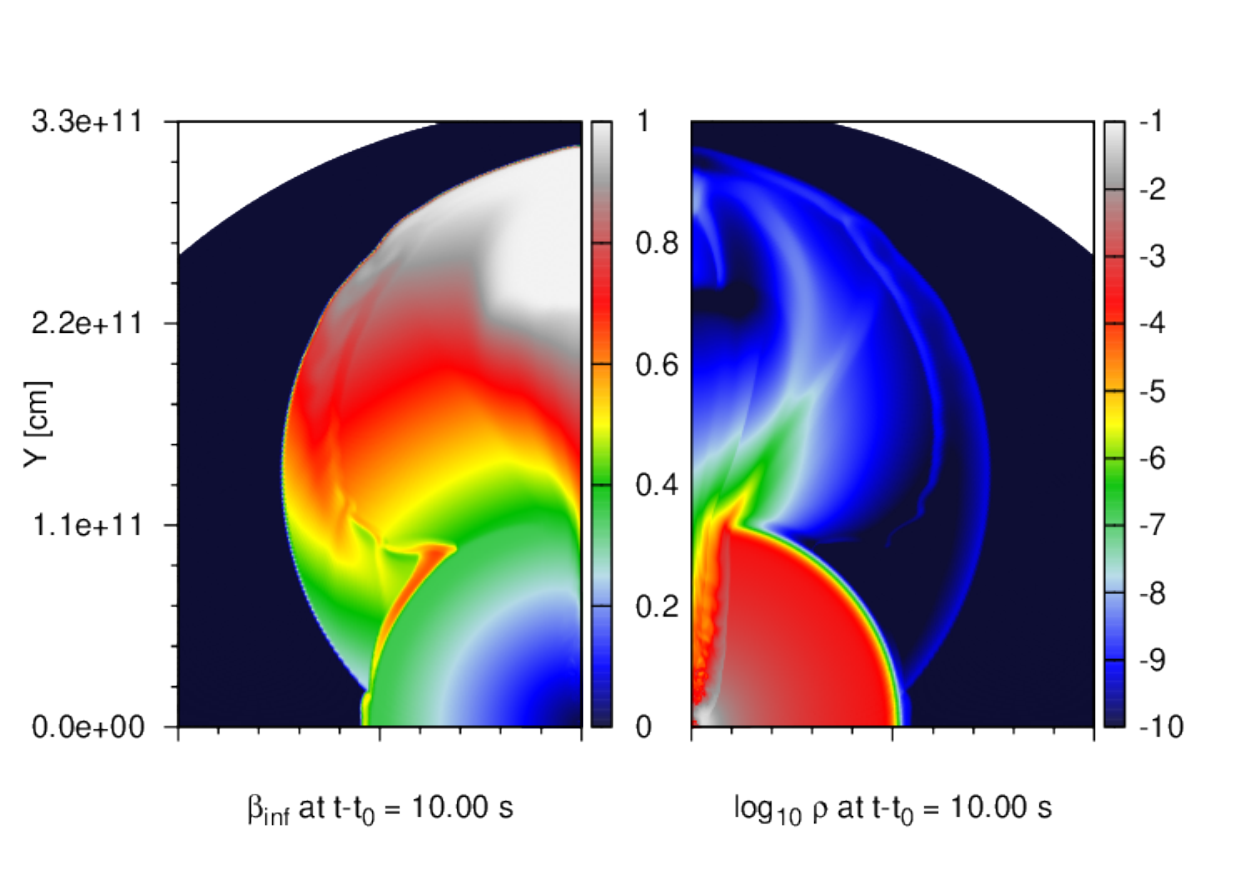}
    %\vspace{4ex}
  \end{subfigure}%% 
  \begin{subfigure}%[b]{0.2\linewidth}
    \centering
    \includegraphics[width=0.49\linewidth]{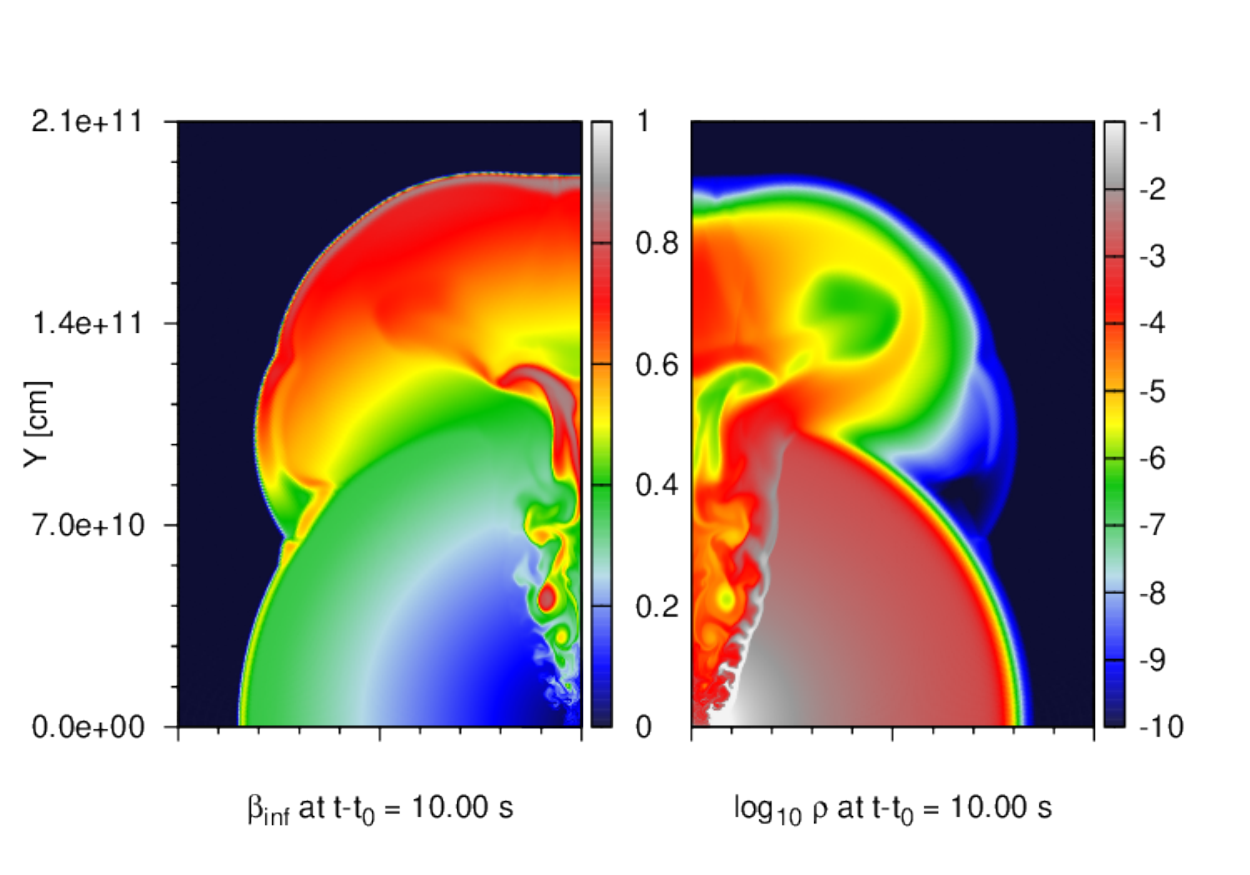} 
    %\vspace{4ex}
  \end{subfigure}
  %\vspace{4ex}
    \begin{subfigure}%[b]{0.2\linewidth}
    \centering
    \includegraphics[width=0.49\linewidth]{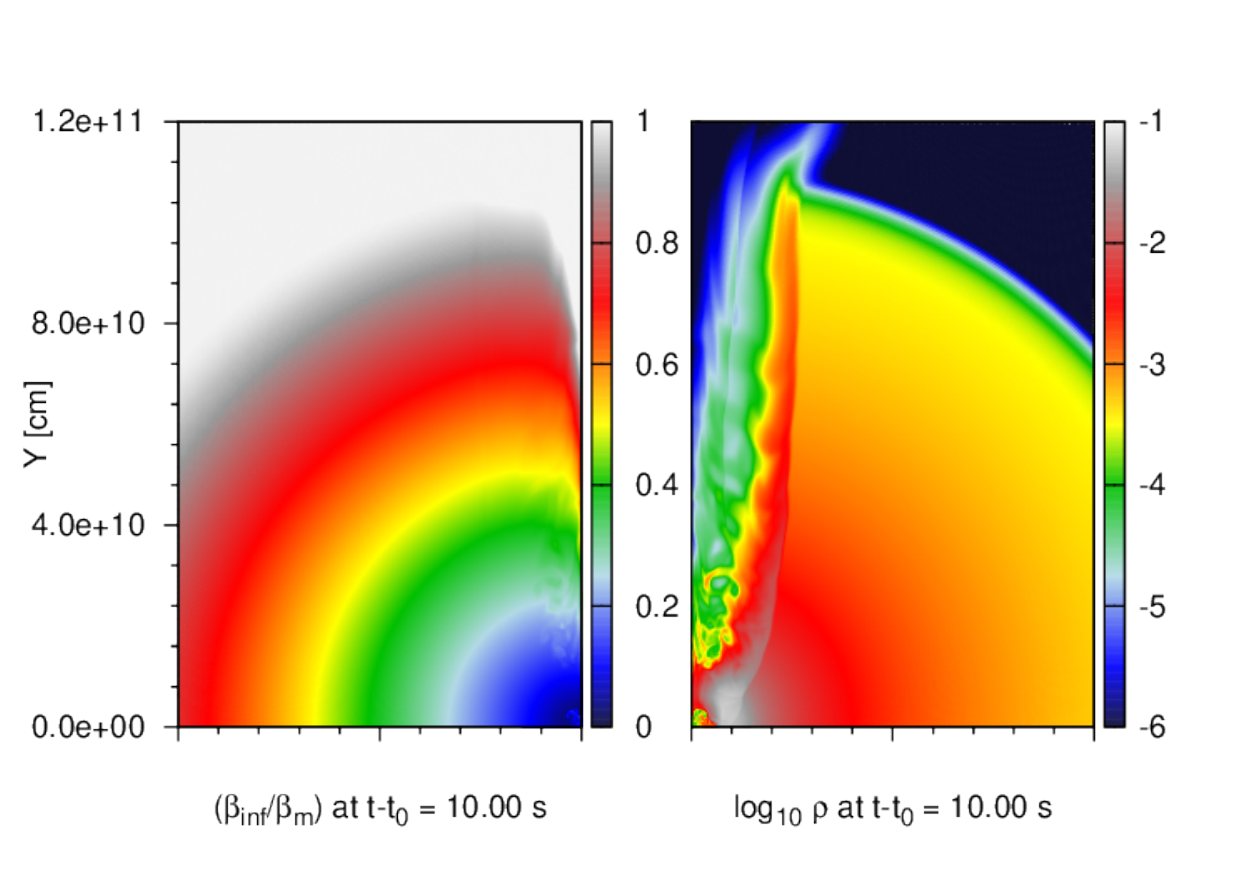}
    %\vspace{4ex}
  \end{subfigure}%% 
    \begin{subfigure}%[b]{0.2\linewidth}
    \centering
    \includegraphics[width=0.49\linewidth]{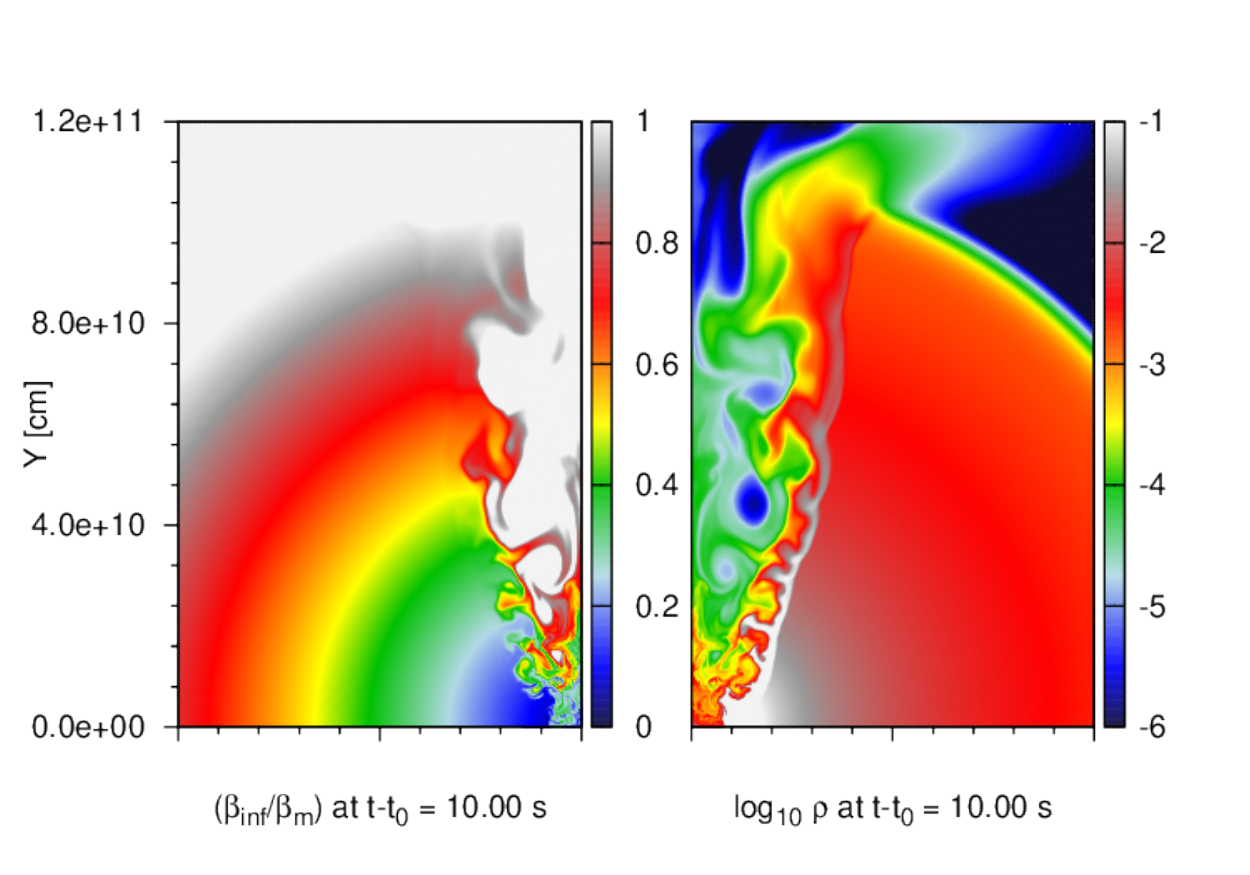} 
    %\vspace{4ex}
  \end{subfigure}
  \caption{
  Velocity and rest-mass density (comoving) maps for the post-breakout cocoon, at the end of simulations ($t-t_0=10$ s).
  Top panels show the whole system, for a focus on the escaped cocoon.
  Bottom panels present a zoom on the trapped cocoon region and its velocity range [the white region (i.e., $\beta_{inf}/\beta_m=1$) illustrates the escaped cocoon].
  The two sub-figures on the left are for the narrow jet case, while the two sub-figures on the right are for the failed jet case (see Table \ref{tab:1}).
  }
  \label{fig:2 sim} 
\end{figure*}

Figure \ref{fig:2 sim} shows maps of the maximum velocity ($\beta_{inf}$) and the rest-mass density ($\rho$; comoving) at the end of the simulation.
Taking into account both of the velocity and the density maps, one can notice that there are two distinct regions of the cocoon: 
\begin{itemize}
    \item The part of the cocoon that is faster than the ejecta ($\beta_{inf} > \beta_m$) and was, or will be able to escape from it into the CSM. We call this part the ``escaped" cocoon.
    \item The part of the cocoon that is not faster than the ejecta ($\beta_{inf} \le \beta_m$) and will not be able to escape from it (assuming that the free expansion phase has been reached, and no further energy exchange would occur in the future). We call this part the ``trapped" cocoon.
\end{itemize}
A close look at the top panels shows that the escaped cocoon is identifiable as having mildly relativistic to sub-relativistic velocities ($\beta \sim 0.35 - 1$).
From the velocity map, it seems that the escaped cocoon expansion is not perfectly homologous (around the on-axis). 
However, it is close enough that we expect that at much later times, fluid elements will eventually be redistributed to make smooth homologous velocity distributions, without significant interactions between the cocoon fluids (see Section \ref{sec:approximations}).
In terms of mass, one can notice that the rest-mass density of this escaped part is much lower than that of the trapped part, especially for the narrow jet case (left).
This feature will be extensively discussed in Section \ref{sec:breakout of the two cc parts}.

Bottom panels of Figure \ref{fig:2 sim} give a zoom on the ejecta (spatially and in terms of velocity range).
This allows one to have a detailed view of the trapped cocoon.
One can notice the ``shocked-jet" cocoon component (i.e., inner cocoon as in \citealt{2011ApJ...740..100B}; low density and high velocity relative to the ejecta), that has not reached the edge of the ejecta yet [the white region in the ($\beta_{inf}/\beta_m$) map].
The size of this component varies widely between the narrow jet model (small) and the failed jet model (large).
This tendency is explained by the relative late breakout time of the failed jet, and the fact that this part is yet to reach its maximum velocity $\beta_{inf}$.
Ideally longer simulation times are better to follow the evolution of this part.
Here, for simplicity, we just assume that this component will eventually breakout without further interactions; hence it is considered as a part of the ``escaped" cocoon, despite spatially being inside the ejecta\footnote{
The only issue with this assumption is that internal energy of the escaped cocoon in the failed jet model, as estimated from numerical simulations, may slightly give overestimated values. 
This is because the internal energy of this part should further be reduced due to sideways adiabatic expansion suffered as soon as it passes the edge of the ejecta (in addition to the radial expansion which can easily be tracked as $\propto t^{-1}$, and has been taken into account).
Still, considering the size of this component, such overestimation would still be within a factor of $\sim 2$.}.

The other noticeable point is the ``shocked ejecta" part of the trapped cocoon (the dense part of the cocoon), and the fact that it is expanding almost homologously.
Therefore, this part, as well as most of the cocoon's mass, is never expected to break out of the ejecta (see Section \ref{sec:cc post tb}; and Figure \ref{fig:keyB}; for more details).

To summarize, the aftermath of the jet propagation through the expanding ejecta (as it is visualized in the bottom panels of Figure \ref{fig:2 sim}) is that, in the frame of the ejecta, most of the ejecta mass in the path of the cocoon is just slightly displaced sideways (trapped), and only a small fraction of it is pushed to the outside (escaped).
This goes against previous claims that jet propagation though the ejecta would simply leave a hole, as most of the cocoon would escape outside of the ejecta.
Hence, it gives a better understanding of jet propagation in the ejecta of BNS mergers (see Section \ref{sec:3}, and in particular Section \ref{sec:why small} for more details).

From an observational point of view, the division of the cocoon, as trapped or escaped, is crucial because only the escaped part is relevant for the cocoon emission (the emission from the trapped cocoon is blocked by the opaque ejecta for an off-axis observer).
The escaped part, with its small mass, diffuses radiation early on ($< 1$ h), and this early period is the only time window for the cocoon emission to be detected.
This is because, at later times, the KN emission from the ejecta shines over the cocoon (due to its much larger mass) [\citealt{2022arXiv221002255H}].
Hence, the trapped cocoon is irrelevant in both time intervals, and should be separated from the escaped cocoon for correct estimation of the cocoon emission.

%%%%%%%%%%%
%%%%%%%%%%%%%%%%%%%%%%%%%%%%%%%%%%%%%%

\subsubsection{Numerical procedure to extract post-breakout cocoon properties}
\label{sec:extraction}
We analyse the simulation data to extract the hydrodynamical properties of the cocoon. 
In particular, we use a post-process algorithm to carefully discriminate between the jet, the ejecta, and the cocoon, at $t=t_1\gg t_b$, as follows:
\begin{itemize}
    \item The jet outflow is identified as fluid elements having
    \begin{eqnarray}
%        \begin{cases}
          \Gamma_{inf} \geqslant 10,\quad
          \text{or}\quad
          \Gamma_{} \geqslant 5.
%        \end{cases}
        \label{eq:jet condition}
    \end{eqnarray}
    These are fiducial minimum values for the jet's Lorentz factor.
    The reason for such values is that, taking into account the relativistic beaming, the opening angle $\theta \sim 1/\Gamma_{}$ is small, and is of the order of typical jet opening angles (assuming that the maximum Lorentz factor will eventually be reached and that the system is ballistic). 
    \item The ejecta material is identified using the following set of conditions: $\beta \leqslant \beta_m$ and $\beta_\theta/\beta_r \sim 0$. The second condition is due to the fact that the ejecta fluid, initially, does not have an angular velocity component, nor it is turbulent.
    \item The CSM is identified as fluid elements that fulfill the followings: $\rho = \rho_{CSM}$, $\beta= 0$, and $r>r_m[\approx c\beta_m(t-t_m)]$.
    \item Finally, the cocoon is identified after subtracting the above components.
    That is, mainly by using the following set of conditions: $\Gamma_{inf}< 10$, $\Gamma_{}< 5$, and $|\beta_\theta| > 0$. Furthermore, the escaped cocoon and the trapped cocoon can be found using $\beta_{inf}>\beta_m$ and $\beta_{inf}\leqslant\beta_m$, respectively.
\end{itemize}

Note that this is a very simplified description (of the algorithm), and several less relevant details have been skipped (e.g., the interaction of the ejecta outer edge with the CSM, etc.).

\subsubsection{Velocity distribution of the cocoon}
\label{sec:v dis}
\begin{figure*}%[ht] 
    %\vspace{4ex}
    \centering
    \includegraphics[width=0.99\linewidth]{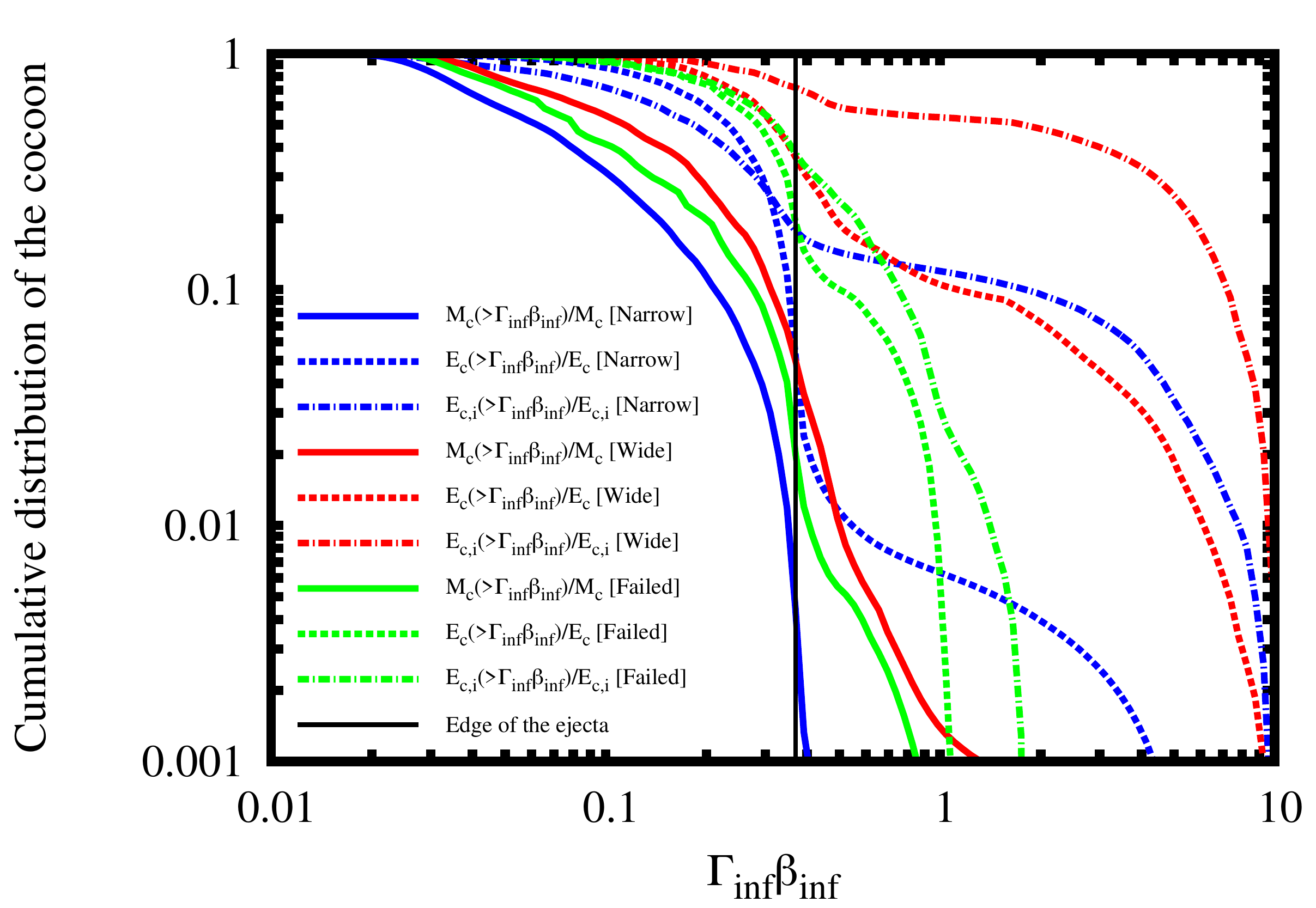}
    %\vspace{4ex}
  \caption{Cumulative distribution of the cocoon's mass (solid), total energy (kinetic + internal; dashed), and internal energy (dashed dotted) for the three simulation models of our subsample (narrow [blue], wide [red] and failed [green]), as a function of the maximum four-velocity. 
  The outer edge of the ejecta ($\beta_m=\sqrt{3/25}$) is shown with a vertical black line.
  Data was taken at the start of the free expansion phase $t-t_0=t_1 \gg t_b$; with $t_1-t_0 = 3$ s for the narrow and wide jet models, and $t_1-t_0=10$ s for the failed jet case (see Section \ref{sec:extraction} for more details).
  }
  \label{fig:3 beta} 
\end{figure*}

Figure \ref{fig:3 beta} shows the cumulative distribution of the cocoon's mass, total energy (kinetic + internal energy) and internal energy as a function of the maximum four-vector ($\Gamma_{inf}\beta_{inf}$).
The mass distribution in the wide and failed jet models is similar, but quite different than that of the narrow jet model.
This is because larger jet opening angles tend to significantly increase the escaped cocoon mass [see Section \ref{sec:3}, in particular equations (\ref{eq:alpha tb}) and (\ref{eq:Mc es}); also see Section \ref{sec:why small}].
Therefore, this trend is due to these two models (wide and failed) having the same jet opening angle (significantly larger than that in the narrow model).

The important point here is that, only a few percent ($< 10\%$) of the cocoon mass ends up being faster than the ejecta edge, i.e., escapes the ejecta.
The overwhelming majority of the cocoon mass ends up being trapped inside the ejecta: $99.6\%$ (narrow), $95\%$ (wide), and $98\%$ (failed).
This is counter-intuitive, especially from the background of collapsar jets and cocoons, and indicates that making parallels between collapsars and BNS merger jets can lead to gross approximations.
The physics behind this tendency will be explored in details in Section \ref{sec:why small}.

For all jet models, the total energy distributions show higher total energy fraction at higher velocities, relative to the mass fractions.  
It also reveals that the total energy fraction of the escaped cocoon [$\sim 5\%$ (narrow), $\sim 36\%$ (wide), and $\sim 19\%$ (failed)] is much higher than its mass fraction.
This is due to the higher velocities of the escaped cocoon, relative to the trapped cocoon.
However, still, in all cases the trapped cocoon energy is largely dominant.

Compared to the total energy distributions, 
the internal energy is not equally distributed throughout the cocoon's energy.
Instead, the higher the velocity of a given cocoon element, the  higher the percentage of its internal energy relative to its total energy.
For instance, the fraction of the cocoon's internal energy escaping the ejecta [$\sim 18\%$ (narrow), $\sim 72\%$ (wide), and $\sim 38\%$ (failed)] is higher than the percentage of the escaping cocoon's total energy.
This trend is due to the contribution of the shocked jet part (of the cocoon) that contains predominantly internal energy, as the result of not being polluted by the ejecta's baryon mass (and kinetic energy) [see Figures \ref{fig:1 sim} and \ref{fig:keyA}; also see Section \ref{sec:pre-breakout sim} for more details on the shocked jet part of the cocoon]. 
This allows this part of the cocoon to theoretically reach very high maximum velocities ($\beta_{inf}$) and to be categorized as a part of the escaped cocoon.
It should be noted that the adiabatic expansion in a freely expanding system implies that the internal energy should continue to drop as $\propto 1/r\propto 1/t$.
However, even so, this trend in the internal energy distribution should stay the same\footnote{Again assuming that the cocoon is freely expanding, and no interactions are taking place between the the different elements, nor with the other components (jet, CSM, etc). 
This is not entirely true due to the short simulation time, the artificially high density of the CSM, etc., but still very reasonable as an approximation.}.

Also, one can notice that the energy (total and internal) distributions in  successful jet models (narrow and wide) extend to mildly relativistic domains, while for the failed jet model, there is a sharp cutoff around $\Gamma_{inf}\beta_{inf} \sim 1-2$.
This is due to the very late breakout time in the failed jet model, caused by the slowly moving jet head (due to 5 times less jet luminosity and 5 times more ejecta mass), resulting in the complete failure of the jet.
This delayed breakout also causes the shocked jet part to be more baryon polluted (by the more baryon rich shocked ejecta part), eventually reducing its maximum velocity.

In summary, here, in the BNS merger case, we show that most of the cocoon (in terms of mass and total energy) ends up being trapped inside the ejecta.
The main cause for this tendency is the expansion of the ejecta (see Figure \ref{fig:keyB} and Section \ref{sec:why small}).
This is a very counter-intuitive result, especially from the background of collapsar jets, where, with the surrounding medium being static in collapsars (stellar envelope), one would naively assume that most of the cocoon is expected to escape.
Therefore, making parallels between the collapsar cocoon, and the BNS merger cocoon, can be misleading and could lead to wrong estimations (and conclusions).
Therefore, this finding goes against several previous studies that have overestimated the cocoon mass when evaluating its emission (\citealt{2017ApJ...834...28N}; \citealt{2018ApJ...855..103P}).
Also, we showed that the escaped part of the cocoon tend to be rich in internal energy.
This is explained by the large contribution of the shocked jet part to the escaped cocoon (for more details see Section \ref{sec:3}; and Figure \ref{fig:keyA} in particular).

%%%%%%%%%%%%%%%%%%%%%%%%
%%%%%%%%%%%%%%%%%%%%%%%%

\subsubsection{Geometry of the escaped cocoon}
\label{sec:geometry}
\begin{figure*}%[ht] 
    %\vspace{4ex}
    \centering
    \includegraphics[width=0.99\linewidth]{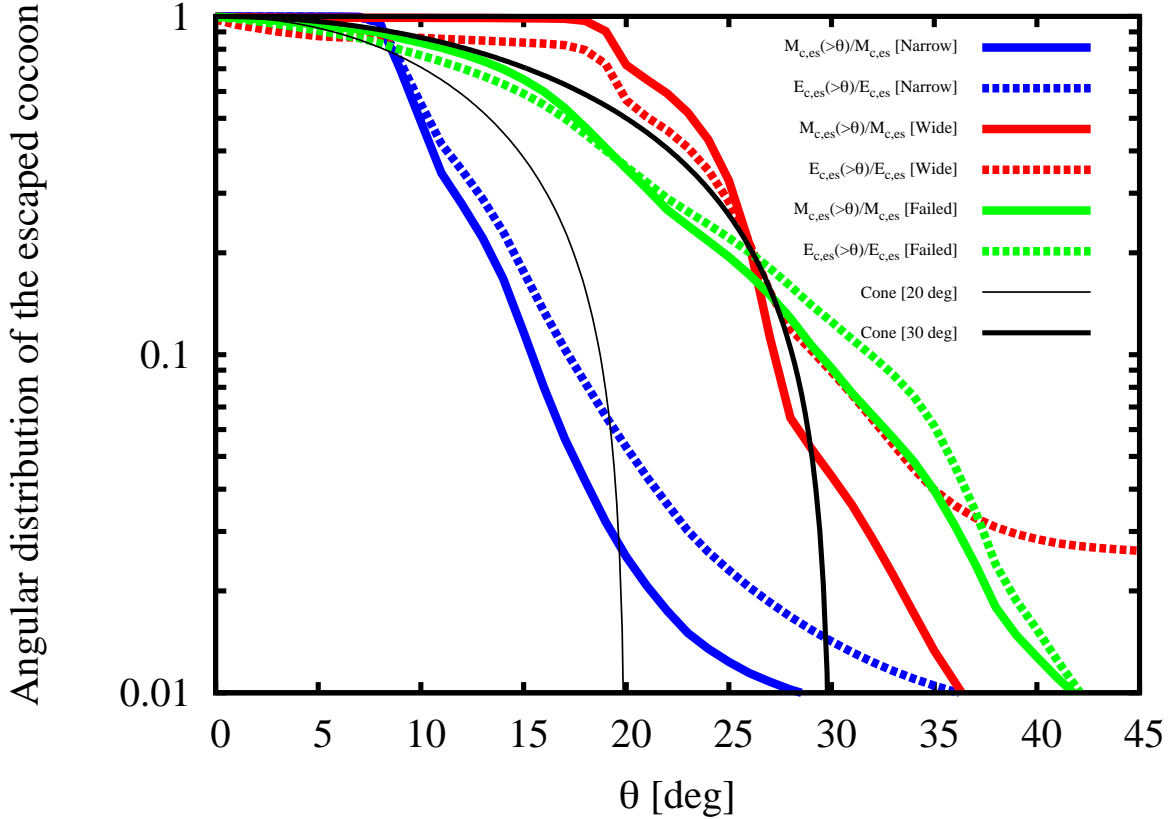} 
  %\vspace{4ex}
  \caption{Cumulative distribution of the escaped cocoon's mass (solid) and total energy (kinetic + internal; dashed), as a function of the angle between the velocity vector (of the expanding escaped cocoon's fluid) and the jet axis; for the narrow (blue), wide (red) and failed (green) jet models.
  The expected cumulative distribution from two homogeneous cones are shown in black: thick for 30 degrees, and thin for 20 degrees.
  Data were taken at the start of the free expansion phase $t-t_0=t_1$, when the cocoon expansion is overall ballistic; with $t_1-t_0 = 3$ s for the narrow and wide jet models, and $t_1-t_0=10$ s for the failed jet case (see Section \ref{sec:extraction}).
  }
  \label{fig:4} 
\end{figure*}

In Figure \ref{fig:4} we show the angular distribution of the escaped cocoon (mass and total energy) for the different jet models.
We find that the escaped cocoon is, overall, well described by a conical structure.
For the narrow jet models, $>90\%$ of the escaped cocoon mass (and total energy) is located within a cone of $\sim 20^\circ$.
For wide and failed jet models, $>90\%$ of the escaped cocoon (mass and energy) is located within a cone of $\sim 30^\circ - 35^\circ$.
This is consistent with \cite{2018MNRAS.473..576G} (see their figure 1).
This is a consequence of the expansion of the ejecta, resulting in $\beta_r \gg \beta_\theta$ in the cocoon [see equation (\ref{eq:b_r gg b_t})], especially true for the escaped cocoon, even after $\beta_{inf}$ is reached. 
This goes against the naive idea that the escaped cocoon takes a spherical shape\footnote{
At much later time, the escaped cocoon-CSM interaction may increase the cone's opening angle, if the CSM density is substantial.}.
This is a very important point to take into account when estimating the optical depth and the cooling emission of the escaped cocoon.

%%%%%%%%%%%

%%%%%%%%%%%%%%%%%%%%%%%%%%%%%%%%%%%%%%

\section{Physical model of the cocoon breakout}
\label{sec:3}

\subsection{Key approximations}
\label{sec:approximations}

Guided by numerical simulation, we take into account the following approximations for our analytic modeling (some of these approximations have already been introduced in Section \ref{sec:2 setup}):
\begin{itemize}

\item We take $\Gamma =1/\sqrt{1-\beta^2}\approx 1$ for the cocoon in the pre-breakout phase $t\leqslant t_b$ (since the maximum velocity of the ejecta is $\beta_m=\sqrt{3/25}\sim0.345$).
Hence, a non-relativistic treatment is acceptable.

\item We use the Bernoulli equation to estimate the maximum velocity of the cocoon [see equations (\ref{eq:Bernoulli Gamma}) and (\ref{eq:Bernoulli Beta})].

\item The initial velocity profile of the ejecta is homologous,
\begin{eqnarray}
r \propto v
\label{eq:r=vt  ejecta}
\end{eqnarray}
This is a very reasonable approximation considering numerical relativity simulations' results (see Section \ref{sec:2 setup}; and Figure 8 in \citealt{2020MNRAS.491.3192H}).

\item For the cocoon, at the breakout time $t_b$, we approximately take that the shape is ellipsoidal, with $r_h/2$ and $r_c$ being the semi-major axis and minor-axis (respectively; see \citealt{2021MNRAS.500..627H}; in particular their Figure 5).
This gives the cocoon's volume as
\begin{eqnarray}
    V_c \approx \frac{4\pi}{3}r_c^2 r_h .
    \label{eq:V_c}
\end{eqnarray}
Furthermore, as shown by numerical simulations, we approximately take that $r_h/2 \gg r_c$ (see Figure \ref{fig:1 sim}).
In other words,
\begin{eqnarray}
    \beta_r\gg \beta_\theta.
    \label{eq:b_r gg b_t}
\end{eqnarray}
This is an ideal approximation for the narrow jet models. 
It is less ideal for wide jet models, but is still very reasonable.

\item At much later times after the breakout time $t\geqslant t_1 \gg t_b$, the escaped cocoon is considered to be freely expanding (i.e., ballistic).
Interaction of the cocoon material with the jet and the ejecta, as well as interaction between cocoon elements themselves, 
can be considered as having an insignificant effect on the  cocoon’s overall dynamics, and hence can be neglected.
This time is found as $t_1-t_0 \approx t_e/(1-\beta_m) \sim 3$ s for successful jet models,
and $t_1-t_0=10$ s for the failed jet model (see Section \ref{sec:timeline}).
This approximation is backed by simulation data showing that variations in the escaped cocoon mass and energy are small after this time.

\item We assume that at much later times after the breakout, $t\gg t_b$ ($t\sim t_1$; see Section \ref{sec:timeline}), every fluid element will eventually converge to its maximum velocity $\beta_{inf}$ [see equations (\ref{eq:Bernoulli Beta}) and (\ref{eq:b_r gg b_t})], neglecting the effect of future interactions.
Also, eventually, a homologous profile for $\beta_{inf}$ is a good approximation,
\begin{eqnarray}
    r(t)\approx c\beta_{inf}(t-t_m) .
    \label{eq:r=vt inf}
\end{eqnarray}

\item During and immediately after the merger, mass is ejected dynamically [by tidal interaction, collision shock, and oscillation of the remnant] (\citealt{2013PhRvD..87b4001H}; \citealt{2013ApJ...773...78B}; \citealt{2015MNRAS.448..541J}; and others).
Post-merger processes also contribute to mass ejection (viscous and neutrino driven winds; \citealt{2018ApJ...860...64F}; \citealt{2020arXiv200700474F}; see \citealt{2019ARNPS..6901918S} for a review).
The ejected mass is not spherically symmetric (e.g., the dynamical part tends to be more concentrated in the equatorial region).
Here, ``ejecta" is used as an umbrella term to refer to the baryon mass through which the jet and the cocoon propagate (and interact), regardless of whether it is bound or unbound; 
in particular, referring to the dynamically expelled mass.
Therefore, in the following, we approximate the density profile of the ``ejecta" to a single power-law function (spherically symmetric) as $\rho_e \propto r^{-n}$ (see footnote \ref{foot:1}).
Furthermore, we only consider the case of a density profile (of the ejecta) with $n=2$.
Hence, $\rho_e \propto r^{-2} \propto v^{-2}$ [see equation (\ref{eq:r=vt  ejecta})].
This is a reasonable approximation considering the properties of the dynamical ejecta in the polar region, where the jet-cocoon structure takes place (see Section \ref{sec:2 setup}; and \citealt{2020MNRAS.491.3192H} for more details, Figure 8 in particular).
Also, this is a helpful simplification as it allows us to find the energy of a (cold) fluid element of the ejecta $dE_e$, with a volume $dV_e$, and a velocity $v=c\beta$, at a given time, as $dE_e \approx \frac{1}{2}\rho_{e}dV_e v_{}^2\propto \beta^{0} dV_e \propto dV_e$ (in the non-relativistic case).
Hence, the energy density throughout the expanding ejecta, at a given time, can be found as
\begin{eqnarray}
    \frac{dE_{e}}{dV_{e}} = \frac{E_{e}}{V_{e}} \propto \text{Const.}  ,
    \label{eq:n=2 case}
\end{eqnarray}
with [using equation (\ref{eq:r=vt inf})]
\begin{eqnarray}
     V_e \approx \frac{4\pi}{3}r_m^3 \approx \frac{4\pi}{3}[c\beta_m(t-t_m)]^3 ,
    \label{eq:V_e}
\end{eqnarray}
and
\begin{eqnarray}
     E_e \approx \frac{1}{6}(c\beta_m)^2 M_e .
    \label{eq:E_e definition}
\end{eqnarray}
In other words, at a given time, the energy density in the ejecta is the same everywhere.

\item Radiation pressure dominates inside the pre-breakout cocoon; 
taking the adiabatic index $\Gamma_A = 4/3$.
Hence, with $\Gamma \sim 1$, at a given time $t\le t_b$, the cocoon's internal energy density can be found as
\begin{eqnarray}
    \frac{dE_{c,i}}{dV_c} \approx 3P_c .
    \label{eq:ei=3pv}
\end{eqnarray}

\item We assume that the internal energy supplied by the jet (into the cocoon), at a given time $t$, is equally distributed within the pre-breakout cocoon.
That is, for a given snapshot, the pressure (i.e., internal energy density) throughout the pre-breakout cocoon is the same everywhere and only depends on time (see Figure \ref{fig:1 sim}).
Hence:
\begin{eqnarray}
    \frac{dE_{c,i}}{dV_c} \approx \frac{E_{c,i}}{V_c} \approx 3P_c \approx \text{Const}.
        \label{eq:P=Const}
\end{eqnarray}

\item Based on \cite{2021MNRAS.500..627H} findings, not all the energy injected by the jet into the cocoon, $E_{in}$, is in the form of internal energy;
around the breakout time, the approximation of equipartition of internal energy $E_{in,i}$ and kinetic energy $E_{in,k}$ gives a reasonable description of numerical simulation data (see top panels of Figure A.1 in \citealt{2021MNRAS.500..627H}),
\begin{eqnarray}
        E_{in}\approx E_{in,i}+E_{in,k}\approx 2E_{c,i},
        \label{eq:equipartition}
\end{eqnarray}
with $E_{in,i}\equiv E_{c,i}$ as the basic source of internal energy in the cocoon.
As explained above, $E_{c,i}$ is equally distributed throughout the cocoon [see equation (\ref{eq:P=Const})].
However, kinetic energy originating from the engine $E_{in,k}$ is actually not equally distributed throughout the cocoon;
numerical simulations show that this component is responsible for creating turbulence, and most importantly, for pushing the shocked ejecta part of the cocoon in the $\beta_{\theta}$ direction creating a dense edge (see density maps in Figure \ref{fig:1 sim}; also see bottom panels in Figure \ref{fig:2 sim}). 
Therefore, although $E_{in,k}$ is not homogeneously distributed throughout the cocoon (${dE_{in,k}}/{dV_c}\neq \text{Const}$), overall this energy is not transferred radially, across different cocoon shells with different velocities [${dE_{in,k}(\beta)}/{dV_c(\beta)}\approx \text{Const}$].
Therefore, considering different shells of the cocoon, each expanding with a radial velocity $\beta$ at a given time, one can find for the jet energy injected into the cocoon,
\begin{eqnarray}
     \frac{dE_{in}(\beta)}{dV_c(\beta)}\approx \frac{E_{in}}{V_c}\approx \text{Const}. ,
     \label{eq:equipartition Ein}
\end{eqnarray}
where
\begin{eqnarray}
     E_{in} = 2L_j (t-t_0)\left[1-\frac{r_h}{c(t-t_0)}\right] ,
     \label{eq:Ein definition}
\end{eqnarray}
and $\frac{r_h}{c(t-t_0)} \approx \langle{\beta_h}\rangle$ is the average velocity of the jet head.
The term $\frac{r_h}{c(t-t_0)}$ accounts for the fraction of the engine energy in the form of unshocked jet, hence it has to be subtracted here (\citealt{2011ApJ...740..100B}). 
And the factor $2$ is to account for the both polar jets ($L_j$ is the luminosity per jet).
\end{itemize}

\subsection{The energy density ratio of the cocoon to the ejecta $\alpha$}
\label{sec:alpha}
Let's consider a cocoon fluid element with a volume $dV_c$, at a given time $t$.
This fluid element was originally an element of the expanding ejecta, and hence it includes a fraction of the ejecta volume $d\tilde{V}_e=dV_c$ (at the same time $t$). 
The total energy of this cocoon element is the sum of the fraction of the kinetic energy of the expanding ejecta ${d\tilde{E}_{e}}$ inside its volume $dV_c$, 
and the fraction of the engine energy that reached this cocoon element ${dE_{in}}$.
Consequently, the energy density of this cocoon fluid element is the sum of the two terms
$\frac{d \tilde E_e}{d \tilde V_e} = 
\frac{d E_e}{d V_e}$ and $\frac{dE_{in}}{dV_{c}}$. 

As explained in Section \ref{sec:approximations}, at a given time $t$, the energy density throughout the ejecta (with a density profile $\rho_e\propto r^{-n}$ and $n=2$, expanding homologously $r\propto v$) can be found having a flat spatial distribution [see equation (\ref{eq:n=2 case})].
Also, the engine energy injected into the cocoon can also be found taking a flat distribution for different cocoon shells [see equation (\ref{eq:equipartition Ein})].
Therefore, the sum for the two above terms, the cocoon's energy density for different shells with different velocities (at a given time), can be found having a flat spatial distribution as well,
\begin{eqnarray}
\frac{dE_c(\beta)}{dV_c(\beta)} = \frac{dE_{e}}{dV_e} + \frac{dE_{in}}{dV_c} \approx \frac{E_{e}}{V_e} + \frac{E_{in}}{V_c} \approx \frac{E_c}{V_c} \approx \text{Const} .
\label{eq:Ec const}
\end{eqnarray}
One can find the total energy of the cocoon as [e.g., using equation (\ref{eq:Ec const})]
\begin{eqnarray}
E_c \approx E_e\left(\frac{V_c}{V_e}\right) + E_{in} . 
\label{eq:Ec two terms}
\end{eqnarray}

In the following, we define the parameter $\alpha$ as the ratio of the cocoon energy density relative to the ejecta energy density, at a given time. 
In other words, the value of $\alpha$ is the value of the energy (kinetic + internal) boost of a given infinitesimal ejecta fluid element after its transformation to an infinitesimal cocoon fluid element.
And since the two energy densities (cocoon and ejecta) do have flat spatial distributions (as a function of $\beta$), so does $\alpha$ [see equations (\ref{eq:n=2 case}) and (\ref{eq:Ec const})]. 
Hence,
\begin{eqnarray}
\alpha = \frac{dE_c/dV_c}{dE_{e}/dV_e} \approx  \frac{E_c V_e}{E_{e}V_c}.
\label{eq:alpha2}
\end{eqnarray}
It can be found as [using equations (\ref{eq:E_e definition}), (\ref{eq:Ein definition}), and (\ref{eq:Ec two terms})]
\begin{eqnarray}
\alpha \approx 1+\frac{12L_j(t-t_0)\left[1-\frac{r_h}{c(t-t_0)}\right]}{(V_c/V_e)\beta_m^2M_{e}c^2},
\label{eq:alpha 1}
\end{eqnarray}
where $V_c/V_e \approx r_hr_c^2/r_m^3$ [see equations (\ref{eq:V_c}) and (\ref{eq:V_e})].
Thanks to the fully analytic formulae of \cite{2021MNRAS.500..627H}, it is possible to find 
both cocoon radii $r_h$ and $r_c$, up to the breakout [follow equation (42) and (30) in \cite{2021MNRAS.500..627H}, respectively; or see equation (\ref{eq:Vc/Ve})].
Hence, using this expression, one can analytically find $\alpha$, at a given time, simply as a function of the parameters of the jet and the ejecta (for more details, specifically at the breakout time $t_b$, see Section \ref{sec:cc pre tb}).

\subsection{The pre-breakout cocoon up to the breakout time $t_b$}
\label{sec:cc pre tb}
As the scope of this study is to understand the cocoon breakout, we will first find the cocoon properties at the breakout time $t=t_b$, just when the cocoon and the jet are about to start escaping the ejecta.

\subsubsection{The parameter $\alpha$ at $t_b$}
At the breakout time,
\begin{eqnarray}
r_h=r_m \equiv r_b \approx c\beta_m t_b.
\label{eq:rb}
\end{eqnarray}
Using equation (\ref{eq:alpha 1}), and $L_{iso,0} \approx 4L_j/\theta_0^2$ (see Table \ref{tab:1}), we can write $\alpha$ at the breakout time as
\begin{eqnarray}
\alpha \approx 1+\frac{3\theta_0^2 L_{iso,0}(t_b-t_0)\left\{1-\beta_m\frac{(t_b-t_m) }{(t_b-t_0)}\right\}}{(V_c/V_e)\beta_m^2M_{e}c^2} .
\label{eq:alpha tb}
\end{eqnarray}

Using the formulation of \cite{2021MNRAS.500..627H} [equations (30) and (34), for $n=2$], 
and $V_c/V_e \approx (r_c/r_b)^2$, we can find that
\begin{eqnarray}
\frac{V_c}{V_e}=\sqrt{\frac{\langle{\eta'}\rangle\langle{\chi}\rangle^2\theta_0^2 L_{iso,0}(t_b-t_0)^3}{8\beta_m^2M_e c^2 (t_b-t_m)^2}} ,
\label{eq:Vc/Ve}
\end{eqnarray}
where, roughly, $\langle{\eta'}\rangle\sim 1/4$ and $\langle{\chi}\rangle \sim 1-2$ [see equations (22) and (28) in \cite{2021MNRAS.500..627H} for their definitions and values].
$t_b-t_m$ is the breakout time, and can easily be found analytically [using equation (44) in \cite{2021MNRAS.500..627H}].
Hence, equation (\ref{eq:alpha tb}) is entirely known, giving a fully analytic expression for $\alpha$, that only depends on the parameters of the jet and the ejecta.

\subsubsection{Mass of the cocoon $M_c$ up to $t_b$}
\label{sec: Mc at tb}
As explained in Section \ref{sec:approximations}, the cocoon shape is best described by an ellipsoid (with $r_h/2$ as the semi-major axis, and $r_c$ as the semi-minor axis; see Figure 5 in \citealt{2021MNRAS.500..627H}).
The equation of this ellipsoid (describing the edge of the cocoon, in one hemisphere) is
\begin{eqnarray}
\left(\frac{y}{r_h/2}\right)^2+ \left(\frac{x}{r_c}\right)^2 =1.
\label{eq:ellipsoid}
\end{eqnarray}
The mass of the cocoon can be found, at a given time $t$, with the following integration,
\begin{eqnarray}
\frac{M_c}{2}=\int_{-r_h/2+r_0}^{r_h/2}dy\pi x^2\rho_{e}(x,y) ,
\label{eq:cc mass at t}
\end{eqnarray}
where $r_0$ is the inner boundary of the cocoon and the jet injection radius in our simulations\footnote{Physically, $r_0$ relates to the fall-back radius (\citealt{2015ApJ...802..119K}; \citealt{2021ApJ...922..185I}), and should be of the same order. 
However, as long as it is much smaller than the initial ejecta radius, its value does not affect the numerical, nor the analytical results (see \citealt{2020MNRAS.491.3192H}; and \citealt{2021MNRAS.500..627H}).}.
The factor $1/2$ accounts for the counter-jet's cocoon in the southern hemisphere.
The initial density of the ejecta is assumed to follow a simple power-law function $\rho_e(r)=\rho_0(r_0/r)^2$, where, with $r_{m,0}=r_m(t_0)$, $\rho_0=\frac{M_{e}}{4\pi r_0^2}\frac{1}{r_{m,0}-r_0}\frac{r_{m,0}}{r_m}\approx \frac{M_{e}}{4\pi r_0^2 r_m}$. 
The cocoon mass can be found as
\begin{eqnarray}
M_c=2\pi \rho_0 r_0^2\int_{-r_h/2+r_0}^{r_h/2} \frac{dy}{1+\left(\frac{y+r_h/2}{x}\right)^2} ,
\label{eq:cc mass at t 2}
\end{eqnarray}
with $0\leqslant x \leqslant r_c$ and $-r_h/2+r_0\leqslant y \leqslant r_h/2$.
However, around the breakout time, as $r_h/2\gg r_c$
and $y+r_h/2\gg x$ [see equation (\ref{eq:b_r gg b_t}); also see Figure \ref{fig:1 sim}], we can write
\begin{eqnarray}
M_c\approx 2\pi \rho_0 r_0^2\int_{-r_h/2+r_0}^{r_h/2} {dy}\left(\frac{x}{y+r_h/2}\right)^2 ,
\label{eq:cc mass at t 3}
\end{eqnarray}
or more simply [using equation (\ref{eq:ellipsoid})],
\begin{eqnarray}
M_c\approx \frac{M_{e}}{2 r_m}\left(\frac{r_c}{r_h/2}\right)^2 \int_{-r_h/2+r_0}^{r_h/2}dy\left[\frac{r_h}{r_h/2+y}-1\right] .
\label{eq:Mc}
\end{eqnarray}
Eventually, with $r_h\gg r_0$, one should find that
\begin{eqnarray}
\frac{M_c}{M_{e}}\approx 2 \left[\ln\left(\frac{r_h}{r_0}\right)-1\right] \left(\frac{r_c}{r_{h}}\right)^2 \frac{r_h}{r_{m}} .
\label{eq:Mc almost final}
\end{eqnarray}
At $t=t_b$, this gives the fraction of the cocoon mass as
\begin{eqnarray}
\frac{M_c}{M_{e}}\approx 2 \left[\ln\left(\frac{r_b}{r_0}\right)-1\right] \frac{V_c}{V_{e}} .
\label{eq:Mc final}
\end{eqnarray}

\subsubsection{Total energy of the cocoon $E_c$ up to $t_b$}
Using the parameter $\alpha$ [see equation (\ref{eq:alpha2})], one can find the total energy of the cocoon simply as
\begin{eqnarray}
E_c \approx \alpha E_e \left(\frac{V_c}{V_e}\right) .
\label{eq:Ec Ee}
\end{eqnarray}

\subsubsection{Internal energy of the cocoon $E_{c,i}$ at $t_b$}
Using equation (\ref{eq:P=Const}), we can find the internal energy of the cocoon as
\begin{eqnarray}
E_{c,i} \approx 3P_c V_c,
\label{eq:Eci}
\end{eqnarray}
where $P_c$ can be found using equation (34) in \cite{2021MNRAS.500..627H}.

\subsection{The post-breakout cocoon: escaped cocoon vs. trapped cocoon}
\label{sec:cc post tb}
Let's consider a given fluid element of the cocoon around the breakout time.
This element will either escape the ejecta, or stay trapped inside it.
As explained in Section \ref{sec:post-breakout sim} [see equation (\ref{eq:Bernoulli Beta})], we approximately consider that each fluid element will eventually converge to its terminal velocity $\beta_{inf}$ at sufficiently later times ($t\gg t_b$).
Therefore, (as mentioned in Section \ref{sec:post-breakout sim}) we consider that the fate of a given cocoon fluid element is solely determined by its final velocity, and whether it is larger or smaller than $\beta_m$ (the maximum velocity of the ejecta),
\begin{equation}
  \beta_{inf} 
    \begin{cases}
      >\beta_m & \text{(Escaped cocoon)} ,\\
      \leqslant \beta_m & \text{(Trapped cocoon)} .
    \end{cases}       
    \label{eq:beta_inf cases}
\end{equation}
In the following, quantities related to the ``escaped cocoon" will be refereed to using the superscript ``$es$".

\subsubsection{Breakout of the two components of the cocoon: Shocked ejecta part and shocked jet part}
\label{sec:breakout of the two cc parts}
As reviewed in Section \ref{sec:pre-breakout sim}, the cocoon is composed of two distinct parts: the ``shocked jet" part (low density and high fraction of internal energy in its total energy) and the ``shocked ejecta" part (much higher density and lower internal energy fraction in its total energy) [see \cite{2011ApJ...740..100B}, the inner and outer cocoons in their Figure 1; also see Figure \ref{fig:1 sim}].
These two part are adjacent to each other, and are constantly being mixed making them difficult to differentiate (see \citealt{2017ApJ...834...28N}; also see \citealt{2021MNRAS.500.3511G} for more details about mixing; see also \citealt{2021MNRAS.503.2499P}).
Hereafter, the ``shocked jet" part is used to refer only to fluid elements of the initially pure shocked jet part that are still able to escaped the ejecta (i.e., with $\beta_{inf}>\beta_m$)\footnote{Here, we categorize the portion of the shocked jet part that is mixed with the shocked ejecta part, ending up being heavily baryon loaded (to the point where it cannot escape the ejecta) as no longer an element of the shocked jet part, but rather as an element of the shocked ejecta part (See Section \ref{sec:fmix}).}. 
In the following, quantities related to the ``shocked jet" part and the ``shocked ejecta" part will be refereed to using the subscript ``$c,j$" and ``$c,e$", respectively.

\begin{figure}%[ht] 
    %\vspace{4ex}
    \centering
    \includegraphics[width=0.99\linewidth]{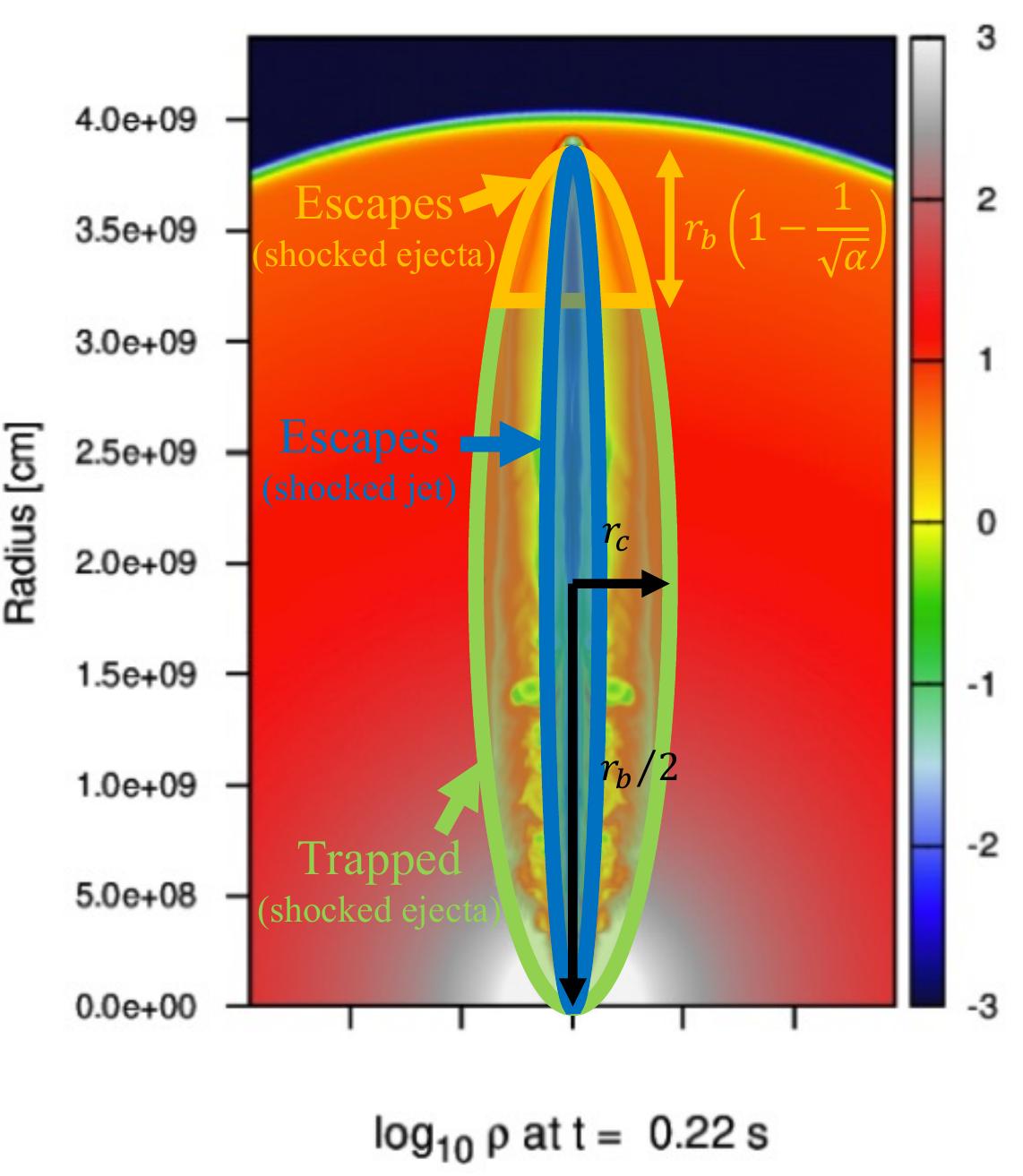} 
  %\vspace{4ex}
  \caption{Density map of the jet, the cocoon, and the surrounding ejecta, at the breakout $t=t_b$, as in our simulation [narrow jet model; same as in the left panel of Figure \ref{fig:1 sim}].
  The two semi-axis of the ellipsoid describing the cocoon's shape are shown (black): $r_b/2$ (half of the breakout radius; major), and $r_c$ (the cocoon width at the breakout time; minor).
  The following parts of the cocoon are highlighted [as found using our analytic model]: i) the escaped part of the shocked ejecta (orange) [approximated to an ellipsoidal cap ($r_b\gg r_c$), with a height $r_b(1-1/\sqrt{\alpha})$; see Figure \ref{fig:keyB}]; ii) the trapped part (green); and iii) the shocked jet component (central ellipsoid; dark blue).}
  \label{fig:keyA} 
\end{figure}

At a given time $t\le t_b$, let's consider that the shocked jet part and the shocked ejecta part have a volume $V_{c,j}$ and $V_{c,e}$, respectively.
On one hand, the cocoon's internal energy is equally distributed throughout $V_c=V_{c,j}+V_{c,e}$ [refer to equation (\ref{eq:P=Const})]; its internal energy density is the same throughout both parts.
On the other hand, the cocoon's (kinetic + internal) total energy density is different (higher in $V_{c,e}$ than in $V_{c,j}$).
This is because of the low mass of the shocked jet part (mass has immigrated to the shocked ejecta part).
The very baryon-poor shocked jet part neighbors the shocked ejecta part. 
Consequently, a fair fraction of the shocked jet part is exposed to baryon loading (e.g., Kelvin–Helmholtz instability), and ends up being dissipated into the shocked ejecta part.
Therefore, in general, the escaping shocked jet part's volume and energy are much smaller compared to the whole cocoon ($V_c \gg V_{c,j}$ and $E_c \gg E_{c,j}$; see Figures \ref{fig:1 sim} and \ref{fig:keyA}); 
allowing one to approximate the energy of the cocoon as equally distributed throughout the shocked ejecta part\footnote{From our simulations, the shocked jet part of the cocoon takes the shape of a very narrow ellipsoid around the jet axis (see Figure \ref{fig:1 sim}; and Figure \ref{fig:keyA} in particular).
Since matter initially located in the shocked jet part has been pushed sideways to the shocked ejecta part, it is safe to consider that the kinetic energy components of the shocked jet part [$(E_{in}/2)(V_{c,j}/V_c)$ (from the engine) and $E_{e}(V_{c,j}/V_e)$ (from the ejecta)] have all immigrated to the shocked ejecta part [see equation (\ref{eq:equipartition})].
Adding the three contributors of energy [engine (kinetic), engine (internal), and ejecta (kinetic)] allows us to write the shocked ejecta's total energy as $E_{c,e} \approx \frac{V_{c,e}}{V_c}\frac{E_{in}}{2}+\frac{V_{c,e}}{V_c-V_{c,j}}\frac{E_{in}}{2} + \frac{V_{c,e}}{V_e}E_e$, and the same expression as in equation (\ref{eq:cc two parts 2/2}) can found after using $V_c\gg V_{c,j}$.
This is due to the shocked jet part being much narrower in width (to the cocoon; see Figure \ref{fig:keyA}): $r_{c,j} \ll r_{c}\Rightarrow V_{c,j}\propto r_{c,j}^2 \ll V_c\propto r_c^2$. 
This is used for the escaped part, where this approximation is particularly good [see equations (\ref{eq:Vc es}) and (\ref{eq:Vcj}) for an evaluation of this approximation].
}.
Hence, the corresponding total energy (kinetic + internal) of each of the two parts of the cocoon can be deduced (respectively) as [refer to equations (\ref{eq:P=Const}) and (\ref{eq:Ec const})]
\begin{equation}
E_{c,j} \approx E_{c,j,i} \approx \left(\frac{V_{c,j}}{V_c}\right)E_{c,i} ,
\label{eq:cc two parts 1/2}
\end{equation}
\begin{equation}
E_{c,e} \approx \left(\frac{V_{c,e}}{V_c}\right)E_{c} ,
\label{eq:cc two parts 2/2}
\end{equation}
and, the corresponding internal energy (respectively) as
\begin{equation}
E_{c,j,i} \approx E_{c,j}   \approx \left(\frac{V_{c,j}}{V_c}\right)E_{c,i},
\label{eq:cc two parts 1/2 2}
\end{equation}
\begin{equation}
E_{c,e,i} = \left(\frac{V_{c,e}}{V_c}\right)E_{c,i} .
\label{eq:cc two parts 2/2 2}
\end{equation}
The fraction of the internal energy for each part can be found as [using equations (\ref{eq:equipartition}), (\ref{eq:Ec two terms}), (\ref{eq:alpha2}), (\ref{eq:cc two parts 2/2}), and (\ref{eq:cc two parts 2/2 2}); and equations (\ref{eq:cc two parts 1/2}) and (\ref{eq:cc two parts 1/2 2}); respectively]
\begin{equation}
\frac{E_{c,e,i}}{E_{c,e}} \approx \frac{E_{c,i}}{E_{c}} \approx \frac{\alpha-1}{2\alpha}, 
\label{eq:fint ejecta part}
\end{equation}
\begin{equation}
\frac{E_{c,j,i}}{E_{c,j}} \sim 1 .
\label{eq:fint jet part}
\end{equation}

%%%%%%%%%%%%%%%%%%%%%%%%

\begin{figure}%[ht] 
    %\vspace{4ex}
    \centering
    \includegraphics[width=0.99\linewidth]{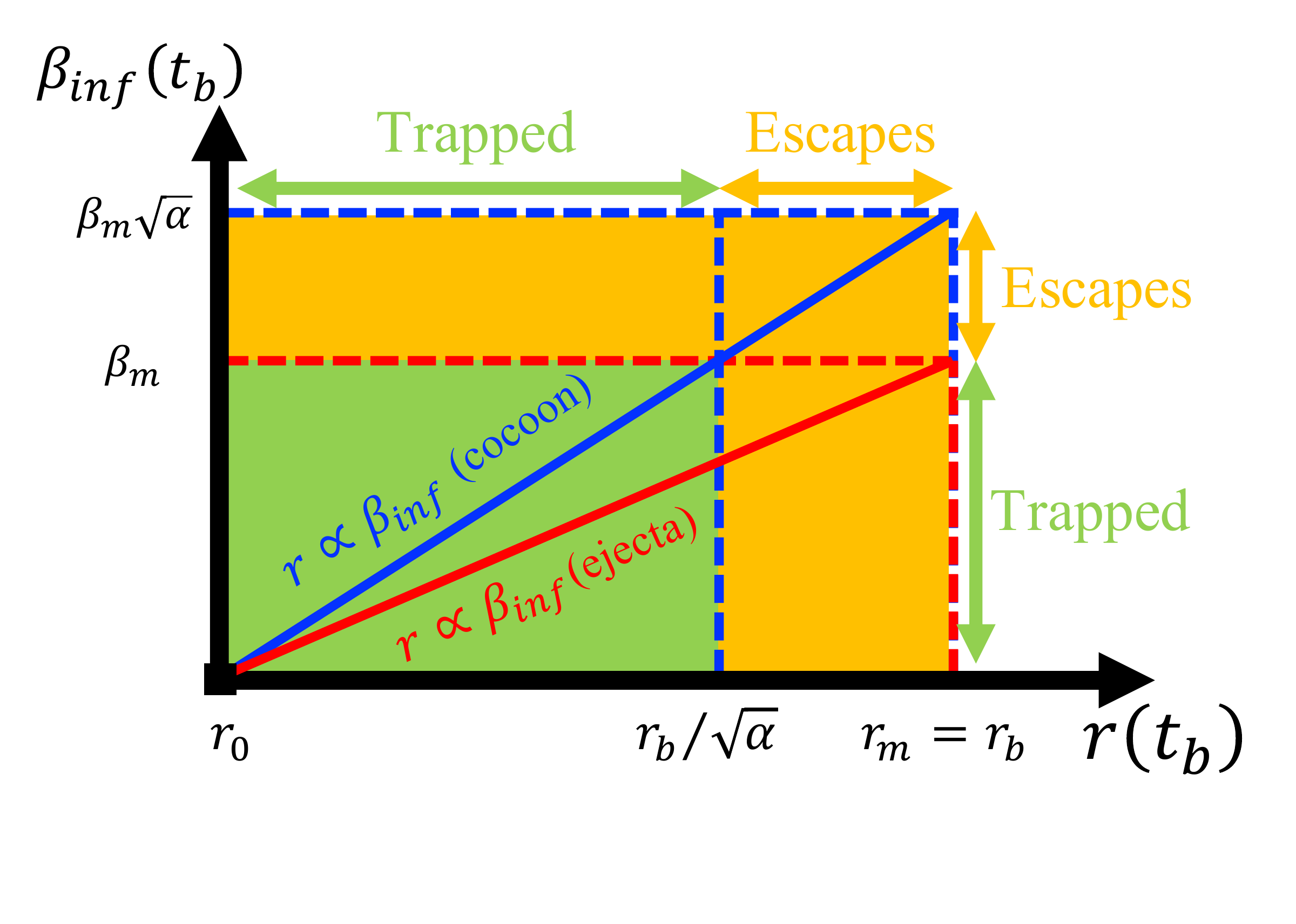} 
  %\vspace{4ex}
  \caption{Velocity profile of the ejecta (red solid line) and of the cocoon (shocked ejecta part; blue solid line).
  The escaped (orange) and the trapped (green) domains of the cocoon (the shocked ejecta part) are identified in terms of radii and maximum velocities ($>\beta_m$ and $\le \beta_m$, respectively), and as a function of the parameter $\alpha$. 
  This was found using the definition of $\alpha$ [equation (\ref{eq:alpha2})], $r\propto \beta_{inf}$, $E_c\propto \beta_{inf}^2$ (non-relativistic case), and the escape criteria [equation (\ref{eq:beta_inf cases})].
  This shows the fiducial case of $\alpha \sim 2$.
  }
  \label{fig:keyB} 
\end{figure}

%%%%%%%%%%%%%%%%%%%%%%%

\subsubsection{Breakout of the shocked ejecta, and the mass of the escaped cocoon $M_c^{es}$}
\label{sec:Mc es}
\paragraph{Energy of the escaped ejecta part $E_{c,e,i}^{es}$ and $E_{c,e}^{es}$:}
\label{sec: Ee es at tb}
As shown in Figure \ref{fig:keyA}, let's consider the cocoon at the moment of the jet breakout $t_b$.
It is worth remembering that the morphology of the cocoon is well described by an ellipsoidal shape [with $r_b/2$ as the semi-major axis, and $r_c$ as the semi-minor axis; also, see equation (\ref{eq:ellipsoid})]. 
As shown in Figure \ref{fig:keyB}, assuming that all cocoon fluid elements will eventually converge to a homologous velocity distribution, $r(t\gg t_b) \propto \beta_{inf}$ [see equation (\ref{eq:r=vt inf})]; 
then, at the breakout time, because the energy has increased by a factor $\alpha$ [see equation (\ref{eq:alpha tb})], the final velocity of the shocked ejecta part has been boosted by $\sqrt{\alpha}$\footnote{The shocked ejecta part of the cocoon is non-relativistic before the breakout and will not reach relativistic velocities due to its high density. Therefore, estimating energy as $E/V\propto \rho v^2$ is reasonable.}.
Using the definition of the escaped cocoon [see equation (\ref{eq:beta_inf cases})], the fate of the shocked ejecta part of the cocoon is decided as follows:
\begin{equation}
   r(t_b)
    \begin{cases}
       > \frac{r_b}{\sqrt{\alpha}} \Rightarrow \beta_{inf} >  \beta_m & \text{(escaped)} ,\\
      \leqslant \frac{r_b}{\sqrt{\alpha}} \Rightarrow \beta_{inf} \leqslant  \beta_m  & \text{(trapped)} .
    \end{cases}       
        \label{eq:escape cases}
\end{equation}
Figure \ref{fig:keyB} shows the division between the escaped/trapped cocoons in the velocity/radial dimensions, as dictated by this condition.
In other words, the escaped shocked ejecta part is located in the outer region of the ellipsoid representing the cocoon.
This region can be approximated to an ellipsoidal cap with a height of $r_b(1-\frac{1}{\sqrt{\alpha}})$, as highlighted in Figure \ref{fig:keyA}.
Hence, the volume of this cap, $V_{c,e}^{es}$, can easily be written as a function of the volume of the ellipsoid (i.e., cocoon $V_c$) and $\alpha$ as\footnote{As a reminder, the volume of the ellipsoidal cap is $V_{cap}=\frac{\pi a b}{3 c^{2}} h^{2}(3 c-h)$, where $a$, $b$, $c$ are the principal semi-axes, and  $h$ is the height of the cap. Here, for the cocoon, $a=b=r_c$, $c=r_b/2$, and $h=r_b(1-1/\sqrt{\alpha})$ [see Figures \ref{fig:keyA} and \ref{fig:keyB}].
With equation (\ref{eq:V_c}) it gives the expression in equation (\ref{eq:Vc es}).}
\begin{eqnarray}
\frac{V_{c,e}^{es}}{V_c}\approx 1+\frac{2}{\alpha^{3/2}}-\frac{3}{\alpha} ,
\label{eq:Vc es}
\end{eqnarray}
and using equations (\ref{eq:P=Const}) and (\ref{eq:Ec const}), one can also find the fraction of the escaped cocoon (in terms of the internal energy and the total energy) as
\begin{eqnarray}
\frac{E_{c,e,i}^{es}}{E_{c,i}}\approx \frac{E_{c,e}^{es}}{E_c}\approx \frac{V_{c,e}^{es}}{V_c}\approx 1+\frac{2}{\alpha^{3/2}}-\frac{3}{\alpha} ,
\label{eq:Ec,e,i,es}
\end{eqnarray}
where $E_{c,e}^{es}$ is the total energy of the escaped shocked ejecta part and $E_{c,e,i}^{es}$ is its internal energy.

\paragraph{Mass of the escaped cocoon $M_c^{es}$:}
\label{sec: Mc es at tb}
As the mass of the shocked jet part is negligible, the mass of the escaped ejecta part of the cocoon is roughly the mass of the cocoon contained in the ellipsoidal cap defined above [see equation (\ref{eq:Ec,e,i,es})]. 
This mass can be calculated with the following integration (see Figures \ref{fig:keyA} and \ref{fig:keyB}):
\begin{eqnarray}
M_c^{es}/2=\int_{\frac{r_b}{2}\left(\frac{1}{\sqrt{\alpha}}-1\right)}^{r_b/2}dy\pi x^2\rho_e(x,y) .
\label{eq:cc cap mass}
\end{eqnarray}
This is the same expression as in Section \ref{sec: Mc at tb} [equation (\ref{eq:cc mass at t})] with the only difference being the lower limit of the integration [in addition to using $r_b\gg r_0$ in equation (\ref{eq:Mc final})].
Replacing gives
\begin{eqnarray}
\frac{M_{c}^{es}}{M_{e}} \approx 2 \left[\frac{1}{\sqrt{\alpha}}-1+\ln\left(\sqrt{\alpha}\right)\right] \left(\frac{V_c}{V_{e}}\right) .
\label{eq:Mc es}
\end{eqnarray}
Using equation (\ref{eq:Mc final}), we can find the fraction of the escaped cocoon as
\begin{eqnarray}
\frac{M_{c}^{es}}{M_{c}}\approx \frac{ \frac{1}{\sqrt{\alpha}}-1+\ln\left(\sqrt{\alpha}\right)}{\ln \left(\frac{r_{b}}{r_{0}}\right)-1} .
\label{eq:Mc es final}
\end{eqnarray}
This is a simple, fully analytical formulation for the escaped cocoon mass; 
with the denominator being the consequence of the ellipsoidal shape of the cocoon [equation (\ref{eq:Mc final})], 
and the numerator being the consequence of the escaped cocoon's location in the outer region of the cocoon where the density $\rho\propto r^{-2}$ is much lower.

\subsubsection{Breakout of the shocked jet cocoon}
\label{sec:shocked jet breakout}
\paragraph{Mixing in the cocoon and the parameter $f_{mix}$:}
\label{sec:fmix}
As previously pointed out, the shocked jet part is very baryon-poor, and its mass density is negligible compared to that of the shocked ejecta part (see Figures \ref{fig:1 sim} and \ref{fig:keyA}) (\citealt{2011ApJ...740..100B}).
However, as the shocked jet part is adjacent to the dense shocked ejecta part, a fraction of the shocked jet part is exposed to being mixed with the shocked ejecta part (\citealt{2017ApJ...834...28N}; also see \citealt{2021MNRAS.503.2499P}).
Such mixing continues to happen throughout the jet propagation, and the shocked jet part of the cocoon is constantly being created (by and near the jet) and dissipated (in the shocked ejecta part, after being mixed with it);
the balance of these two processes determines its final properties (at $t_b$).
Therefore, the size of the shocked jet part of the cocoon, at the breakout time, is closely related to the degree of mixing between the two parts of the cocoon (\citealt{2011ApJ...740..100B}; \citealt{2013ApJ...777..162M}; \citealt{2017ApJ...834...28N}; \citealt{2018MNRAS.477.2128H}; \citealt{2021MNRAS.500.3511G}; \citealt{2021MNRAS.503.2499P};
etc.).
However, the physics of such mixing is not well understood, even with numerical simulation, as it is dependent on resolution, magnetic field, dimensionality, etc. (also jet instability \citealt{2019MNRAS.490.4271M}; although it is most relevant in the collapsar case \citealt{2021MNRAS.500.3511G}).

Since the energy of the shocked jet part of the cocoon is mostly composed of internal energy, the expected maximum velocity of this part is high (high enthalpy).
Therefore, if no mixing happens, the shocked jet part of the cocoon is expected to escape the ejecta (see Figure \ref{fig:keyA}).
This trend implies that the internal energy fraction (internal energy/total energy; not counting the rest mass energy) in the escaped cocoon should be much higher than the average cocoon.
Therefore, the degree of mixing (of the shocked jet part with the shocked ejecta part) is closely correlated to the internal energy fraction of the escaped cocoon relative to the average internal energy fraction of the entire cocoon,
i.e., the higher the mixing degree is, the more comparable the internal energy fraction of the escaped cocoon to that of the entire cocoon is, and vice versa. 
Interestingly, since the internal energy of the escaped cocoon determines the cocoon's luminosity, the degree of mixing is closely related to this observable quantity (\citealt{2017ApJ...834...28N}).

Here we introduce a new parameter, $f_{mix}$, that reflects the degree of mixing of the shocked jet part of the escaped cocoon with the shocked ejecta part, and that could be measured from numerical simulation. 
We define $f_{mix}$ as the ratio of the internal energy fraction of the cocoon over the internal energy fraction of the escaped cocoon,\footnote{Later after the cocoon break out, shock heating in the (artificially dense) CSM does convert some of the cocoon's kinetic energy to internal energy.
This internal energy is not (and should not be) included in the definition of $f_{mix}$. 
As such, the measurement of internal energy of the escaped cocoon from our simulations at later times may falsely indicate lower values for $f_{mix}$ due to such heating in the escaped cocoon.
Therefore, it is better to define it at the moment of the breakout $t_b$.}
\begin{eqnarray}
{f_{mix}} = \frac{E_{c,i}/E_{c}}{E_{c,i}^{es}/E_{c}^{es}} ,
\label{eq:fmix}
\end{eqnarray}
where $0< f_{mix}\leqslant 1 $.
In the limit that the shocked jet part is fully mixed with the shocked ejecta part, this parameter will take the value of 1.
In the limit where the shocked jet part is poorly mixed with the shocked ejecta part, this parameter is expected to be much smaller than 1:
\begin{equation}
  f_{mix} 
    \begin{cases}
      \sim 1 & \text{(Fully mixed)} ,\\
      \ll 1 & \text{(Poorly mixed)} .
    \end{cases}       
    \label{eq: fmix cases}
\end{equation}
From our numerical simulations, we find this parameter to take values in the range $\sim \frac{1}{2}-\frac{2}{3}$.
Therefore, for our analytic model, we set this parameter as
\begin{eqnarray}
  f_{mix} \sim {2}/{3}  .
\label{eq:fmix value}
\end{eqnarray}

It is worth stressing that, as mixing is very sensitive to parameters such as resolution, dimensionality, breakout time, etc., it is entirely understandable if our mixing values differ from values found from other studies.
Therefore, we caution that this value should not be regarded as a universal face value, but rather as an informed guess based on our numerical simulations' setup.

\paragraph{Internal energy of the escaped cocoon $E_{c,i}^{es}$:}
\label{sec: Eci es}
It is worth recalling that the shocked jet part is basically composed of internal energy, and that it all belongs to the escaped cocoon ($E_{c,j} = E_{c,j,i}= E_{c,j}^{es}= E_{c,j,i}^{es}$ and $V_{c,j} \equiv V_{c,j}^{es}$). 
Also, the escaped ejecta part of the cocoon has the same internal energy fraction as the entire shocked ejecta part of the cocoon [see equation (\ref{eq:fint ejecta part})],
\begin{eqnarray}
\frac{E_{c,e,i}^{es}}{E_{c,e}^{es}} \approx \frac{E_{c,e,i}}{E_{c,e}} \approx \frac{\alpha-1}{2\alpha} .
\label{eq:es e fint}
\end{eqnarray}
Using the definition of $f_{mix}$ [equation (\ref{eq:fmix})], the fraction of internal energy in the escaped cocoon can be found as
\begin{eqnarray}
\frac{E_{c,i}^{es}}{E_{c}^{es}} \approx \frac{1}{f_{mix}} \frac{E_{c,i}}{E_{c}} .
\label{eq:es fint}
\end{eqnarray}
Using $E_c^{es} = E_{c,e}^{es}+E_{c,j}^{es}$ and $E_{c,i}^{es} = E_{c,e,i}^{es}+E_{c,j,i}^{es}$ (remembering that $E^{es}_{c,j,i}\approx E^{es}_{c,j}$), and plugging equation (\ref{eq:es e fint}) allows us to find a relationship between the internal energies of the two escaped parts (at $t=t_b$),
\begin{eqnarray}
\frac{E_{c,j,i}^{es}}{E_{c,i}} \approx \frac{2\alpha(1/f_{mix}-1)}{2\alpha-(\alpha-1)/f_{mix}} \frac{E_{c,e,i}^{es}}{E_{c,i}} ,
\label{eq:Ecj i es}
\end{eqnarray}
and volumes as [using equations (\ref{eq:cc two parts 1/2 2}) and (\ref{eq:Ec,e,i,es})]
\begin{eqnarray}
\frac{V_{c,j}^{es}}{V_c} \approx \frac{2\alpha(1/f_{mix}-1)}{2\alpha-(\alpha-1)/f_{mix}}  \frac{V_{c,e}^{es}}{V_c}.
\label{eq:Vcj}
\end{eqnarray}
The total energy of the escaping shocked jet part of the cocoon (as a function of the escaping shocked ejecta part) can be deduced as [using equations (\ref{eq:es e fint}) and (\ref{eq:Ecj i es})]
\begin{eqnarray}
\frac{E_{c,j}^{es}}{E_c} = \frac{(1/f_{mix}-1)(\alpha - 1)}{2\alpha-(\alpha-1)/f_{mix}}\frac{E_{c,e}^{es}}{E_c} .
\label{eq:Ecj}
\end{eqnarray}
Then, the fraction of the escaped cocoon, in terms of the cocoon total energy, can be found by adding the term $\frac{E_{c,e}^{es}}{E_c}$ to each side of equation (\ref{eq:Ecj}) and plugging equations (\ref{eq:Ec,e,i,es}) as
\begin{eqnarray}
\frac{E_{c}^{es}}{E_c} = \left(\frac{1+\alpha}{2\alpha-(\alpha-1)/f_{mix}}\right)\left(1+\frac{2}{\alpha^{3/2}}-\frac{3}{\alpha}\right) ,
\label{eq:Ec es tot}
\end{eqnarray}
and in terms of internal energy as [plugging equation (\ref{eq:es fint}) in equation (\ref{eq:Ec es tot})]
\begin{eqnarray}
\frac{E_{c,i}^{es}}{E_{c,i}} =\frac{1}{f_{mix}}  \left(\frac{1+\alpha}{2\alpha-(\alpha-1)/f_{mix}}\right)\left(1+\frac{2}{\alpha^{3/2}}-\frac{3}{\alpha}\right) .
\label{eq:Eci es tot}
\end{eqnarray}

And so, we can analytically evaluate the mass and energy composition of the escaped cocoon with simple equations [equations (\ref{eq:Mc es final}), (\ref{eq:Ec es tot}), and (\ref{eq:Eci es tot}); see equation (\ref{eq:alpha tb}) for $\alpha$].
Their results and accuracy are discussed in Section \ref{sec:4}.

\subsection{Cocoon growth after the breakout time}
\label{sec: fg}
We follow the cocoon evolution in numerical simulations, at much later times after the breakout time, later than the engine's active timescale,
until the cocoon can be considered as freely expanding ($t\sim t_1\gg t_b$; $t_1 \sim 3-10$ s; see Section \ref{sec:timeline}).

Comparison with simulations shows that the above analytic model (Section \ref{sec:cc post tb}) is consistent with numerical simulations at describing the cocoon breakout around the breakout time.
However, simulations show that the cocoon's growth (escaped and trapped) is not over after the breakout.
We notice that the [total and internal (after accounting for adiabatic expansion)] energy continues to grow in all models as long as the engine is active.
This growth is observed for both the escaped and the trapped cocoon.

This is logical, as the jet-cocoon/ejecta-cocoon interaction are not entirely over;
in particular, i) the continued expansion of the cocoon throughout the ejecta, and ii) the non-zero mixing of the jet's fluid with the cocoon's fluid inside the ejecta as the jet injection continues until $t_e -t_0 =2$ s (near the collimation shock in particular)\footnote{In particular, around the radial inner edge of the jet, as the engine is turned off, mixing is enhanced.}.
These two types of interactions are thought to be independent, and hence, should ideally be characterised separately.
However, based on our simulations (for more details see Table \ref{tab:2}), we found that this late time growth can roughly be characterized by one parameter.
This parameter, $f_g$, is defined as the growth of the cocoon (escaped and trapped) in energy and baryon mass, from $t_b$ to $t_1$.
Hence, for the entire cocoon we can get
\begin{eqnarray}
E_{c} (t_1) \approx& E_{c}(t_b)\times f_g ,\\
M_{c} (t_1) \approx& M_{c}(t_b)\times f_g ,\\
E_{c,i} (t_1) \approx& E_{c,i}(t_b)\left[\frac{t_b}{t_1}\right]\times f_g ,
\label{eq:Ec fg}
\end{eqnarray}
and particularly for the escaped cocoon we can also get
\begin{eqnarray}
E_{c}^{es} (t_1) \approx& E_{c}^{es}(t_b)\times f_g ,\\
M_{c}^{es} (t_1) \approx& M_{c}^{es}(t_b)\times f_g ,\\
E_{c,i}^{es} (t_1) \approx& E_{c,i}^{es}(t_b)\left[\frac{t_b}{t_1}\right]\times f_g .
\label{eq:Ec es fg}
\end{eqnarray}
Based on our numerical simulation, we find that overall, $f_g \sim 2$ captures reasonably well this late time growth (for reference, analytic and simulations results are presented at Table \ref{tab:2} for a quantitative comparison).
It is worth mentioning that the fractions of the escaped cocoon (mass, energy, and internal energy), relative to the whole cocoon, at the breakout time remain the same at $t_1$.
This is because they are all independent of this parameter ($f_g$ is canceled out).

Hence, the system of analytic equation modeling the cocoon breakout in NS mergers is closed.
Ultimately, the combination of the analytic model presented here, with the model in \cite{2021MNRAS.500..627H}, allows one to find the properties of the escaped cocoon (mass, total energy, and internal energy; see Table \ref{tab:2}) as a function of following six parameters of the ejecta: mass $M_e$, density profile's power-law index $n$ ($=2$ here), and the maximum velocity $\beta_m$; 
and the jet: isotropic equivalent luminosity $L_{iso,0}$, opening angle $\theta_0$, and  delay time between the merger and the jet launch $t_0-t_m$.

\section{Discussion}
\label{sec:4} 
\subsection{Comparison with simulations}
\label{sec:Comparison with simulations}
As of the time of writing, the analytic modeling presented here was carried out with one goal in mind:
in order to analytically estimate the cocoon's emission, as an EM counterpart, for future GW170817-like events --
as a function of the parameters of the ejecta and the central engine (in particular, in the LIGO O4 era; see Section \ref{sec:1}).
Therefore, considering the theoretical and observational uncertainties, it is reasonable to estimate the cocoon's properties within a factor of a few.

In Table \ref{tab:2}, we list the breakout times, $\alpha$, the cocoon's (escaped + trapped), and the escaped cocoon's properties [mass, energy (kinetic + internal), and internal energy], at $t_1\gg t_b$, as measured from our numerical simulations 
(see Section \ref{sec:extraction} for more details on simulations' data).
Analytic values (using our model in Section \ref{sec:3}) are also presented for comparison.

In our model, the parameter $\alpha$ [calculated using equation (\ref{eq:alpha tb})] is crucial.
Table \ref{tab:2} shows that our analytic model does capture the value of $\alpha$ in simulations reasonably well.
This results in a good consistency of our analytic results with our simulations, especially for the successful jet case (narrow and wide) [in most cases errors are within $\sim 20-30\%$;
and in a few exceptions the difference is by a factor $\sim 2$, which is still very reasonable considering the goal of this study].
We stress that this is the first time that the cocoon breakout has been modeled to such extent, and with such success.

\begin{figure*}%[ht] 
    %\vspace{4ex}
  \begin{subfigure}%[b]{0.2\linewidth}
    \centering
    \includegraphics[width=0.99\linewidth]{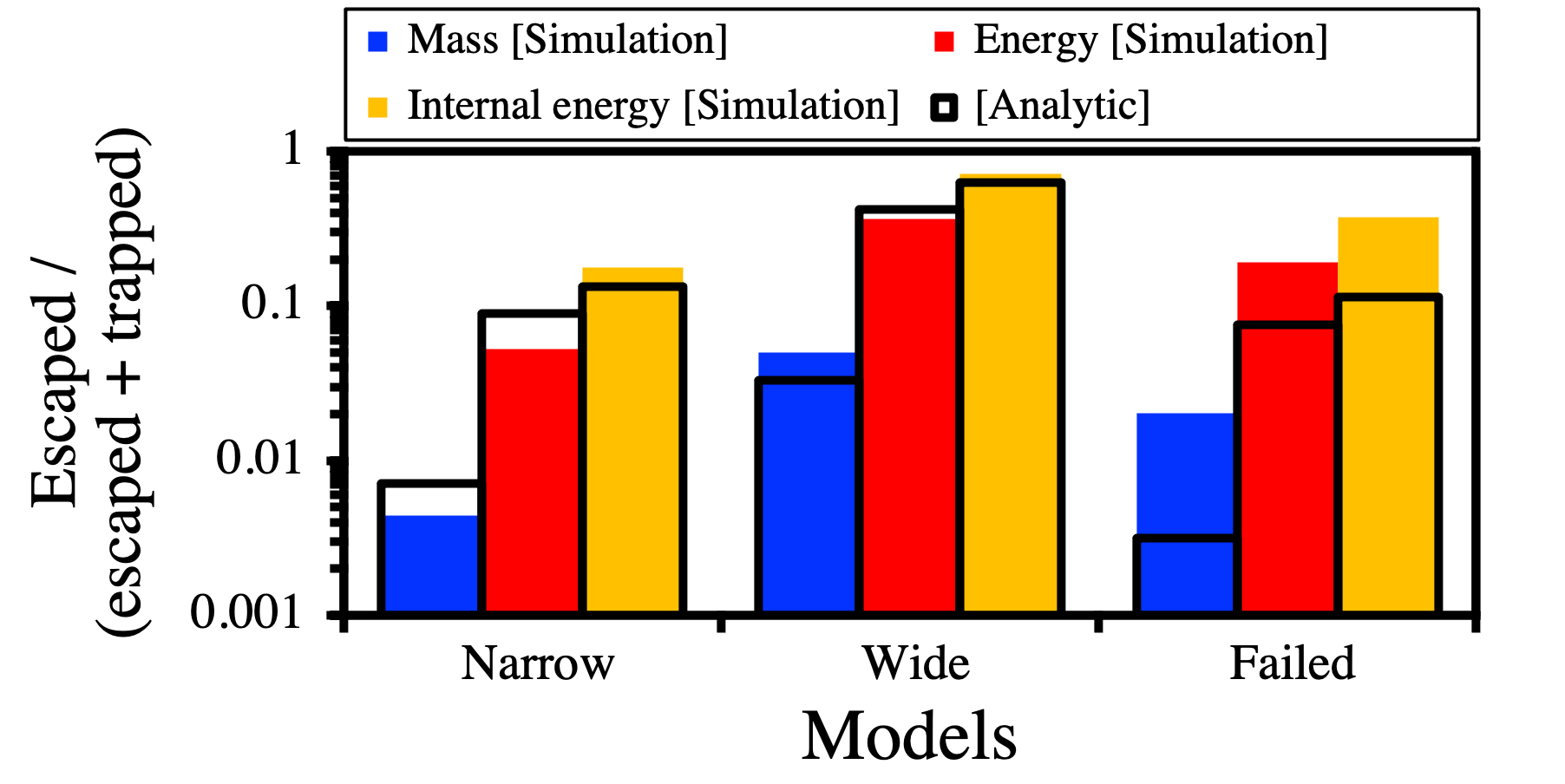}
    %\vspace{4ex}
  \end{subfigure}%% 
  %\vspace{4ex}
  \caption{Mass (blue), energy (red), and internal energy (orange) fractions of the escaped cocoon as measured from simulations (in the laboratory frame), for the narrow, wide, and failed jet models, at the free expansion phase ($t_1$).
  Histograms with black borders indicates the fractions found using our fully analytic model [see equations (\ref{eq:Mc es final}), (\ref{eq:Ec es tot}) and (\ref{eq:Eci es tot}) in Section \ref{sec:3}, respectively]. The actual values can be found in Table \ref{tab:2}.}
  \label{fig:data} 
\end{figure*}

Our analytic model is less accurate at capturing the properties of the cocoon for the failed jet model (compared to successful jet models), with differences up to a factor $\sim 4-5$ compared with simulations.
In particular, the cocoon's total mass and internal energy seem to suffer the most.
We suspect that this is caused by the contribution of two major factors.
First, the analytic model for jet propagation in \cite{2021MNRAS.500..627H} has been calibrated with successful jet models only\footnote{Throughout this study we used the calibration coefficient $N_s=0.46$, 
as measured in \cite{2020MNRAS.491.3192H} and \cite{2021MNRAS.500..627H}, for successful jet models.
It is not clear whether this value is appropriate for the failed jet model.
The parameter $\langle{\eta'}\rangle$ was properly measured from simulations giving: $\langle{\eta'}\rangle=0.25$ for successful jet models [see Figure 1 and equation (22) in \citealt{2021MNRAS.500..627H}]; and $\langle{\eta'}\rangle=0.3$ for the failed jet model.
}; 
and considering the much longer jet breakout timescale in the failed jet model, this might have contributed to errors in the analytical values of $r_c$ (and $r_b$).
Second, the assumption of an ellipsoidal cocoon shape (and $\beta_r \gg \beta_\theta$) works quite well in the successful jet case, 
where the jet head motion is quite fast compared to the sideways expansion (see Figure 5 in \citealt{2021MNRAS.500..627H}). 
However, this might not be a good approximation for the failed jet, where the cocoon shape (as observed with our simulations) is, sort of, hybrid between ellipsoidal and conical (see the right panel in Figure \ref{fig:1 sim}; also see footnote \ref{foot:cone})\footnote{It should be noted that this could be linked to the tendency of 2D simulations to produce a dense polar plug (\citealt{2003ApJ...586..356Z}; \citealt{2010ApJ...717..239L}; \citealt{2013ApJ...777..162M}; \citealt{2018MNRAS.479..588G}; etc.).
}.
This explains why our analytic estimation of the failed cocoon's escaped mass (using an ellipsoidal mass) is much lower than in the numerical simulation (see Figure \ref{fig:data}).
Nevertheless, considering the purpose of our analytic modeling, we consider that these analytic values are still reasonably good, and hence we consider that our analytic modeling is successful.

Figure \ref{fig:data} shows, in a much visual form, the fraction of the escaped cocoon, out of the entire cocoon, [again, in terms of mass, total energy (kinetic and internal, not including the rest mass energy), and internal energy].
The analytic model seems to capture reasonably well the complex process of cocoon breakout (again, especially for successfully jet models), as all values are consistent with simulations, within a factor of a few.
Furthermore, from Figure \ref{fig:data}, analytical modeling agrees with simulations, confirming that the overwhelming majority of the cocoon mass is in the trapped cocoon, never escaping the ejecta, and only a small fraction can escape; while in terms in energy, a larger fraction of the cocoon total energy (although, still, less than half) escapes (see Section \ref{sec:v dis} and Figure \ref{fig:3 beta}; for a deep analysis see Section \ref{sec:why small}).
Also, the behavior of the fractions of the escaped internal energy of the cocoon, showing the opposite trend (compared to mass and total energy fractions), by tending to take larger values, can also be confirmed (see Figure \ref{fig:3 beta}).
As discussed in Section \ref{sec:v dis}, this is due to the nature of the shocked jet part of the cocoon, and its ability to achieve high velocities, hence escaping, with its rich internal energy composition.

%%%%%%%%%%%%%%%%%%%%%%%%%%%%%%%%%%
\begin{table*}
\caption{
A comparison of the results found using the analytic model presented in Section \ref{sec:3}, compared with numerical simulations (in the laboratory frame; see Section \ref{sec:extraction}), at the free expansion phase ($t_1$). 
The breakout time $t_b-t_0$ and $\alpha$ have been calculated using equation (44) in (\citealt{2021MNRAS.500..627H}) and equation (\ref{eq:alpha tb}), respectively.
The cocoon's mass, total energy, and internal energy have been calculated using equations (\ref{eq:Mc final}), (\ref{eq:Ec Ee}), and (\ref{eq:Eci}), respectively.
The escaped cocoon's mass, total energy (not including the rest-mass), and internal energy have been calculated using equations (\ref{eq:Mc es final}), (\ref{eq:Ec es tot}) and (\ref{eq:Eci es tot}), respectively.
}
\label{tab:2}
\begin{tabular}{ll|ll|lll|lll}
  \hline
              &  & $t_b-t_0$ & $\alpha$ & $M_c$ & $E_c$ & $E_{c,i}$ & $M_c^{es}$ & $E_c^{es}$ & $E_{c,i}^{es}$ \\
        Model &  & [s]       &               & [$M_\odot$] & [erg] & [erg] & [$M_\odot$] & [erg] & [erg] \\
        \hline
    Narrow & Simulation & $0.222$	&$1.54$	&$3.0\times 10^{-4}$  &$4.0 \times10^{48}$	&$2.0 \times10^{46}$	&$1.4\times 10^{-6}$	&$2.1\times 10^{47}$	&$3.6\times 10^{45}$	\\
    & Analytic & $0.203$	&$1.47$	&$1.4\times 10^{-4}$ &$1.8\times 10^{48}$	&$4.1\times 10^{46}$	&$9.9\times 10^{-7}$	&$1.6\times 10^{47}$	&$5.5\times 10^{45}$	\\
         \hline
    Wide & Simulation &$0.412$ &$2.14$	&$1.1\times 10^{-3}$ &$2.4\times 10^{49}$	&$7.3\times 10^{47}$		&$3.6\times 10^{-5}$ &$8.8\times 10^{48}$	&$5.2\times 10^{47}$	\\
    & Analytic &$0.450$	&$2.63$	&$1.0 \times10^{-3}$  &$2.0\times 10^{49}$	&$1.1\times 10^{48}$ &$3.5\times10^{-5}$ &$8.2 \times 10^{48}$ &$6.9\times10^{47}$ \\
        \hline
    Failed & Simulation &$2.590$	&$1.52$	&$2.5\times 10^{-3}$	&$5.4\times 10^{49}$	&$7.3\times 10^{47}$	 &$5.0\times 10^{-5}$ &$1.0\times 10^{49}$	&$2.7\times 10^{47}$ \\
    & Analytic &$2.974$	&$1.42$ &$9.7\times 10^{-3}$ &$6.3\times 10^{49}$	&$2.7\times 10^{48}$ 	&$3.0\times 10^{-5}$ &$4.8\times 10^{48}$	&$3.1\times 10^{47}$ \\
  \hline
 \end{tabular}
\end{table*}

%%%%%%%%%%%
%%%%%%%%%%%%%%%%%%%%%%%%%%%%%%%%%%%%%%

\subsection{Why the mass of the escaped cocoon is so small?}
\label{sec:why small}
Our analytic model is consistent with simulation, in the sense that the overwhelming majority of the cocoon mass cannot 
escape from the ejecta ($\sim 90 \% - 99\%$).
This is a counter-intuitive result.
In the following we present a breakdown of the physics behind this fact.

The most important parameter is $\alpha$ [see equation (\ref{eq:alpha tb}) for its definition]. 
Let's consider a fiducial case of $\alpha \sim 2$\footnote{There is a limit on how large $\alpha$ can be in the \textit{s}GRBs scenario; it is very difficult to have values of $\alpha$ much larger than $2$ considering the parameter-space of the jet and the ejecta [see equation (\ref{eq:alpha tb})].}.
By definition, this means that the energy density inside the cocoon has been boosted by a factor of 2 [see equation (\ref{eq:alpha2})].
As $E_c\propto \beta^2$, this implies a boost in the cocoon's velocity ($\beta_{inf}$) by about $\sqrt{2}\sim 1.4$, considering that this energy is spread equally (see Section \ref{sec:approximations}).
As the expansion is homologous, the region of the escaping cocoon [i.e, exceeding the velocity of the edge of the ejecta; see equation (\ref{eq:beta_inf cases})] is situated in $({r}/{r_b})>1/{\sqrt{\alpha}}\sim 0.7$ [i.e., in the outer $\sim 30\%$ radii of the cocoon; see equation (\ref{eq:escape cases}) and Figures \ref{fig:keyA} and \ref{fig:keyB}].
Recalling that the shape of the cocoon is ellipsoidal, the volume of this shape is a smaller fraction of the cocoon volume [$\sim 20\%$ for $\alpha\sim 2$; see equation (\ref{eq:Ec,e,i,es})]\footnote{\label{foot:cone}In the limit where the cocoon shape is close to conical (rather than ellipsoidal), the escaped shocked ejecta part of the cocoon will occupy a much larger volume fraction, $({V_{c,e}^{es}}/{V_c}) \approx 1-\alpha^{-3/2}\sim 0.4-0.5$ [compared to the ellipsoidal cap in the ellipsoidal cocoon case; see equation (\ref{eq:Vc es}); giving (${V_{c,e}^{es}}/{V_c})\sim 0.1-0.2$].
The conical shape also dramatically increases the mass of the escaped cocoon [to $M_c^{es}/M_c \approx 1-\alpha^{-1/2}\sim 0.2$], and hence, it partially explains the high escaped cocoon mass measured for the failed jet model (see Figure \ref{fig:data}).}.
This roughly corresponds to the total energy fractions of the escaping cocoon (see Figures \ref{fig:3 beta} and \ref{fig:data}).

Concerning the escaped mass fraction, there are two other critical points whose combination reduce the mass of the escaped cocoon significantly.
Firstly, the density profile scaling as $\rho_c \sim \rho_e \propto r^{-2}$.
Secondly, the ellipsoidal geometry of the cocoon.
These two imply that: 
i) there is a large contribution of the inner dense regions in the cocoon, and a small contribution of the outer less dense regions (e.g., compare to the case of a conical geometry). 
In fact, the cocoon is denser than the ejecta on average [by about a factor $\sim 5$; see equation (\ref{eq:Mc final}); with $r_b/r_0\sim 4$].
In other words, the cocoon is weighted by the inner part.
And, ii) the location of the escaped cocoon being in the outer ellipsoidal cap [$r\sim (0.7-1) r_b$], which is the least dense region of the cocoon [$\sim 2$ times less dense than the ejecta on average; see equations (\ref{eq:Vc es}) and (\ref{eq:Mc es})].
Hence, (i) and (ii) together suppress the escaped cocoon mass relative to the cocoon mass (by suppressing its relative density) by a factor of $\sim 5\times 2\sim 10$ (for $\alpha \sim 2$).
Note that, if the cocoon geometry were conical, then (i) would have no effect, and (ii) would still have suppressed the escaped cocoon mass by about the same factor  ($\sim 2$); contributing by, at least, a factor $\sim 5$ difference compared to the ellipsoidal case (the volume of the escaped cocoon would have also been boosted further increasing the escaped mass; for more information see footnote \ref{foot:cone}).

Hence, in summary, the combination of these two effects, as discussed above [volume of the polar cap of an ellipsoidal shape ($\sim \frac{1}{5}$), and the density profile combined with the ellipsoidal geometry ($\sim \frac{1}{2}\times \frac{1}{5}$)], explains why the fraction of the escaped cocoon mass should be in the order of a few $\%$ ($\sim 2\% $ for $\alpha \sim 2$).

\subsection{Discussion}
\subsubsection{Previous works}

There has been several numerical studies on the cocoon emission (\citealt{2018MNRAS.473..576G}), its contribution to the KN (\citealt{2021MNRAS.502..865K}; \citealt{2021MNRAS.500.1772N}), and other (see Section \ref{sec:1}).

\cite{2017ApJ...834...28N} was the first to extensively study the phenomenon of cocoon breakout;
they presented modeling of the collapsar cocoon and estimated its emission for the first time.
Overall, \cite{2017ApJ...834...28N} presented a very extensive (and comprehensive) modeling of the cocoon breakout and its hydrodynamical properties, and gave the first one-zone estimation of the post-breakout cocoon properties.
Our work here (simulations in particular) indicates that several of the ideas introduced by \cite{2017ApJ...834...28N} (and \citealt{2011ApJ...740..100B}), in the context of collapsars, are still true (and hence, very useful) in the context of NS mergers as well;
in particular concerning the shocked jet and shocked ejecta (stellar medium in \citealt{2017ApJ...834...28N}), and their mixing [see Figures \ref{fig:1 sim} and \ref{fig:keyA}].
\cite{2017ApJ...834...28N} did also present an attempt to analytically model the cocoon (its emission in particular) in the context of \textit{s}GRBs, mainly by directly applying their collapsar model (see Section 6 in \citealt{2017ApJ...834...28N}).

Our model, here, has been developed by constantly relaying on numerical simulations.
To our best knowledge, this is the first analytic modeling of the cocoon, exclusively in the context of \textit{s}GRBs, in terms of its breakout and late time evolution.
Our model's main result is that modeling of the cocoon breakout in the context of \textit{s}GRBs, where the medium is expanding, is trickier than it is in collapsars.

\subsubsection{Implications on previous estimations of the cocoon's cooling emission}
Previously, several studies estimated the cocoon's emission in \textit{s}GRBs (e.g., \citealt{2017ApJ...834...28N}; \citealt{2018ApJ...855..103P}).
Here, we found, for the first time, that the mass of the escaped cocoon (the relevant part for cocoon emission) should be in the order of $\sim 10^{-5} - 10^{-6} M_\odot$ (see Table \ref{tab:2}), with a very steep density profile (roughly as $\propto  r^{-8}$; \citealt{2022arXiv221002255H}; also, notice the steep mass decline in Figure \ref{fig:3 beta});
the rest of the cocoon mass ($\sim 10^{-4} - 10^{-3} M_{\odot}$) is trapped inside the ejecta.
Consequently, we question previous estimation of the cocoon's cooling emission (in NS mergers/\textit{s}GRB's context);
as previously assumed cocoon masses seem to have been overestimated by orders of magnitude [e.g., \citealt{2017ApJ...834...28N} plugging equation (3) in Section 6; and equation (7) in \citealt{2018ApJ...855..103P}].
Internal energy of the cocoon, critical to its emission, seems to have also been over simplified [e.g., equation (12) in \citealt{2018ApJ...855..103P}];
in reality the cocoon of \textit{s}GRBs contains much less internal energy fraction than the cocoon of \textit{long} GRBs (see Figure 1 in \citealt{2021MNRAS.500..627H}).
Also, another overlooked aspect (of the escaped cocoon) in previous analytical studies is its angular distribution, clearly conical (see Figure \ref{fig:4}), instead of the assumed spherical distribution  (\citealt{2017ApJ...834...28N}; \citealt{2018ApJ...855..103P}).

These much lower cocoon masses have many implications on the cocoon emission;
they imply that the r-process powered emission is dim (compared to the emission from jet shock heating) making the total timescale of the emission from the escaped cocoon shorter (in the order of seconds to tens of minutes, before the KN emission dominates over);
also this implies that the photosphere is smaller and temperature is higher, resulting in an increase in the observed magnitude (i.e., fainter) at optical and UV bands, at a given time [taking $\kappa \sim 1$ cm$^2$ g$^{-1}$ \citealt{2020ApJ...901...29B} and \citealt{2022arXiv220406861B}; e.g., see right panel of Figure 3 in \citealt{2018MNRAS.473..576G}].
The trapped cocoon is enclosed by the ejecta and the escaped cocoon, and is effectively unobservable with the KN in the background.
This would make it much more challenging to detect the cocoon emission (see \citealt{2022arXiv221002255H} for more details).
Also, at early times, considering that, in successful jets, the escaped cocoon can attain mildly relativistic velocities, a proper relativistic treatment would be necessary to estimate the cocoon emission.

In summary, future estimation of the cocoon emission in \textit{s}GRBs should take into account the escaped part of the cocoon.

\subsubsection{Future applications}
Our model is fully analytical and reasonably accurate (within a factor of a few).
The fact that it is only dependent on the parameters of the jet and the ejecta makes it a powerful tool.
For instance, it can be used to explore a wide parameter space of the jet and the ejecta, and evaluate the cocoon properties, and its emission, accordingly.
This can be applied to GW170817-like events to model early EM counterparts and understand their environment (the jet and the ejecta properties).
This can also be used to investigate some unidentified/poorly understood transients [e.g., X-Ray Flashes (XRFs); see \citealt{2002ApJ...571L..31Y}; \citealt{2022arXiv221002255H}; and Hamidani \& Ioka in preparation].

\subsubsection{Limits of our analytic modeling}
\label{sec:limits}
Our modeling has its limits, as the setup presented here has been simplified.
First, it is limited to the case of $n=2$ ($\rho \propto r^{-n}$).
This is consistent with the density profile of the inner part of the dynamical ejecta in NS mergers, in particular in the polar region where the jet-cocoon structure takes place,
but overlooks the outermost part ($n\sim 2.5-3.5$).
Second, recent numerical relativity simulations have indicated the presence of a fast tail component of the ejecta (\citealt{2014MNRAS.437L...6K}; \citealt{2017PhRvD..96h4060K}; \citealt{2018ApJ...867...95H}; \citealt{2018ApJ...869..130R}).
This component is still not very well understood, and depends on the type of EOS in particular (\citealt{2017PhRvD..96h4060K}; \citealt{2018ApJ...869..130R}). 
Studies have shown that its mass is much smaller than that of the ejecta [e.g., see Table 2 in \citealt{2018ApJ...869..130R}].
At first glance, considering our results, the fast tail mass could be of a comparable order of magnitude to that of the escaped cocoon; 
roughly the fast tail mass is $\sim 10^{-5} M_{\odot}$ [although in extreme cases (the EOS and the mass ratio) it could be up to  $\sim 10^{-4} M_{\odot}$ (\citealt{2017PhRvD..96h4060K}; \citealt{2018ApJ...869..130R})].
Taking into account that the fast tail is more concentrated in the equatorial plane, than in the polar region, where the escaped cocoon is supposed to propagate (see Figure \ref{fig:4}; and Figure 8 in \citealt{2020MNRAS.491.3192H}), this would reduce its effective mass (mass of the fast tail facing the escaping cocoon) by roughly two orders of magnitude [especially in the narrow jet case ($\sim 20^\circ$; see Figure \ref{fig:4})].

Hence, mass and total energy of the cocoon dominates over the fast tail (in the polar region).
Still, such fast tail mass could affect the escaped cocoon (especially the shocked jet part), and could transfer some of the relativistic jet energy to the escaped cocoon.
More importantly, shock heating in the fast tail is a very important effect, as it would boost the escaped cocoon's internal energy, hence making its emission brighter.

It has been shown that 2D simulations produce a numerical artifact in the form of a polar plug (\citealt{2003ApJ...586..356Z}; \citealt{2010ApJ...717..239L}; \citealt{2013ApJ...777..162M}; and \citealt{2018MNRAS.479..588G}).
This has been shown to affect the jet propagation.
Ideally, 3D simulations are better for solving the jet propagation.
However, 3D simulations have their downside; resolution in 3D simulations would be much lower than in 2D simulations (for the same computation power).
This would affect mixing in the jet and the cocoon, etc.
Considering the focus of our study (being on the cocoon), we choose to relay on 2D simulations.
This is justified, as 3D simulations would not dramatically affect the overall cocoon's properties (Figure B1 in \citealt{2018MNRAS.479..588G}), and especially considering our goal (estimation within a factor of a few). 
Hence, future studies should consider the effect of 3D simulations, as well as other effects not included here (such as the effect of the magnetic field; especially on mixing and jet stability: \citealt{2019MNRAS.490.4271M}; \citealt{2021MNRAS.500.3511G}).

\section{Summary and Conclusion}
\label{sec:5}
We presented numerical simulations of the cocoon breakout in NS mergers, for three different types of jets: narrow, wide, and failed (see Figures \ref{fig:1 sim}, \ref{fig:2 sim}; and Table \ref{tab:1}).
We followed the cocoon evolution for timescales much longer than the breakout time (up to $\sim 10$ s $\gg t_b-t_0$).
In particular, we analysed the distribution of mass and energy in the cocoon (see Figure \ref{fig:3 beta}), finding that, contrary to previous considerations, only a tiny fraction of the cocoon manages to escape from the ejecta ($\sim 0.5-5 \%$ in terms of mass).
Also, we analyzed the angular distribution of the cocoon after the breakout, finding that it takes a conical ($\sim 20^\circ-30^\circ$; rather than spherical) distribution (see Figure \ref{fig:4}) as a result of the expansion of the ejecta.

In addition, we presented a fully analytic model of the cocoon breakout (see Section \ref{sec:3}).
As illustrated in Figures \ref{fig:keyA} and \ref{fig:keyB}, our model allows us to analytically estimate the mass, kinetic energy, and internal energy of the cocoons (escaped and trapped), using the ``the energy boost factor by the jet" $\alpha$ [see equation (\ref{eq:alpha2})] -- all as a function of the parameters of the ejecta [mass $M_e$, density profile's power-law index $n$ ($=2$ here), and maximum velocity $\beta_m$] and the jet [isotropic equivalent luminosity $L_{iso,0}$, opening angle $\theta_0$, and the delay between the merger time and the jet launch time $t_0-t_m$].
We showed that our analytic modeling is in a good agreement with simulations for all jet models, within a factor of a few [see Figure \ref{fig:data}; and Table \ref{tab:2}].

To our best knowledge this is the first time that the cocoon breakout in NS mergers has been understood and modeled to such depth.
We argue that our model is a very useful tool, in particular at estimating the properties of the escaped cocoon, and hence its cooling emission.
This could be applied to future GW170817-like events, and could reveal new details on NS mergers and their environments, provided that early localization and follow-up observations will be achieved (\citealt{2022arXiv221002255H}).

Interestingly, our analytic model reveals that the cocoon of NS mergers' jets is inherently inefficient at escaping the ejecta, especially in terms of mass.
Even though the cocoon is accelerated by the jet, the surrounding ejecta is expanding fast enough for most of the cocoon to stay trapped within it,
making a striking contrast to the collapsar case (see Section \ref{sec:why small}). 
As a result, the order of the fraction of the escaped cocoon mass should be in the order of a few percent (see Figure \ref{fig:data}).
The other dominant cocoon is trapped and hidden (unobservable) inside the ejecta.
To our best knowledge, this is the first time that this finding has been revealed.
Previous estimations of the cocoon emission in \textit{s}GRBs (e.g., \citealt{2017ApJ...834...28N}; \citealt{2018ApJ...855..103P}), as well as future estimations, should take this into account.

%%%%%%%%%%%%%%%%%%%%%%%%%%%%%%%%%%
%%%%%%%%%%%%%%%%%%%%%%%%%%%%%%%%%%

\section*{Acknowledgements}
\addcontentsline{toc}{section}{Acknowledgements}
We thank
Amir Levinson, 
Banerjee Smaranika, 
Bing Zhang,
Bing Theodore Zhang,
Kazumi Kashiyama, 
Kazuya Takahashi, 
Kenta Kiuchi, 
Kohta Murase, 
Koutarou Kyutoku, 
Kyohei Kawaguchi
Masaomi Tanaka, 
Masaru Shibata, 
Pawan Kumar,
Shota Kisaka,
Shuta Tanaka,
Suzuki Akihiro,
Tomoki Wada, 
Tsvi Piran, 
Wataru Ishizaki,
and, Yudai Suwa,
for fruitful discussions and comments. 
    
We thank the participants and the organizers of the workshops with the identification number YITP-T-19-04, YITP-W-18-11 and YITP-T-18-06, for their generous support and helpful comments. 
    
Numerical computations were achieved thanks to the following: Cray XC50 of the Center for Computational Astrophysics at the National Astronomical Observatory of Japan, and Cray XC40 at the Yukawa Institute Computer Facility.
   
This work was partly supported by JSPS KAKENHI nos. 20H01901, 20H01904, 20H00158, 18H01215, 17H06357, 17H06362, 22H00130 (KI). 

\section{Data availability}
The data underlying this article will be shared on reasonable request to the corresponding author.

%%%%%%%%%%%%%%%%%%%%%%%%%%%%%%%%%%%%%%%%%%%%%%%%%%



\bibliographystyle{mnras}
\bibliography{04.1-mnras} % if your bibtex file is called ?example.bib

%%%%%%%%%%%%%%%%%%%%%%%%%%%%%%%%%%%%%%%%%%%%%%%%%%

%%%%%%%%%%%%%%%%% APPENDICES %%%%%%%%%%%%%%%%%%%%%
\appendix

% Don't change these lines
\bsp	% typesetting comment
\label{lastpage}
\end{document}